\documentclass[5p,sort&compress]{elsarticle}
\usepackage{mathpazo}
\usepackage{booktabs}
\usepackage{multirow}
\usepackage{color}

\usepackage{amssymb,amsmath}
\usepackage{bm}
\usepackage{sansmath}
\usepackage{mathrsfs}
\usepackage{url}
\usepackage{upgreek}
\usepackage{graphics}
\usepackage[switch]{lineno}

\journal{Elsevier}

\begin{document}

\begin{frontmatter}



\title{Atmospheric effects in astroparticle physics experiments and the challenge of ever greater precision in measurements}


\author[label1]{Karim Louedec}
\ead{karim.louedec@in2p3.fr}

\address[label1]{Laboratoire de Physique Subatomique et de Cosmologie (LPSC), Universit\'e Grenoble-Alpes, CNRS/IN2P3, 38026 Grenoble cedex, France}

\begin{abstract}
Astroparticle physics and cosmology allow us to scan the universe through multiple messengers. It is the combination of these probes that improves our understanding of the universe, both in its composition and its dynamics. Unlike other areas in science, research in astroparticle physics has a real originality in detection techniques, in infrastructure locations, and in the observed physical phenomenon that is not created directly by humans. It is these features that make the minimisation of statistical and systematic errors a perpetual challenge. In all these projects, the environment is turned into a detector medium or a target. The atmosphere is probably the environment component the most common in astroparticle physics and requires a continuous monitoring of its properties to minimise as much as possible the systematic uncertainties associated. This paper introduces the different atmospheric effects to take into account in astroparticle physics measurements and provides a non-exhaustive list of techniques and instruments to monitor the different elements composing the atmosphere. A discussion on the close link between astroparticle physics and Earth sciences ends this paper.
\end{abstract}


\end{frontmatter}


\section{Introduction}
\label{intro}
Recent years have seen the development of major infrastructure around the Earth in order to increase considerably the performances of experiments in astroparticle physics and cosmology. Unlike other fields in science where measurements are made on a physical phenomenon created in laboratory, research in astroparticle physics has originality in detection techniques and infrastructure locations. Experiments are operated over large desert areas as the Cherenkov Telescope Array (CTA)~\cite{CTA_web}, the Pierre Auger Observatory~\cite{PAO_web} or very soon the LSST telescope~\cite{LSST_web}, in oceans or ice with ANTARES~\cite{ANTARES} and IceCube~\cite{IceCube}, respectively, or even in space with projects as the AMS-02 experiment~\cite{AMS} or soon the JEM-EUSO telescope~\cite{JEM-EUSO}. In all these projects, and more than any other experience in subatomic physics, minimising statistical and systematic errors is a challenge because the physical phenomenon observed is not produced by man himself: scientists are just observers. Thus, scientists build ever larger detectors to go further in the knowledge. However, owning a large detector is not a necessary and sufficient requirement to push the limits of our knowledge: the systematic error still lurks and demands from scientists an excellent understanding of their detector. The temptation to increase the duty cycle of the detector in order to reduce still more the statistical error should not obscure the need to control the associated increase in systematic error. Therefore there is a point where these two errors become inseparable and where the optimisation of detector performance can be compared to the concept of \emph{yin} and \emph{yang}.

In all these projects, the environment is turned into a detector medium or a target. The atmosphere is probably the environment component the most common in astroparticle physics, usually used as a giant calorimeter in cosmic ray experiments or as an irreducible detection volume in the case of ground-based astrophysics surveys. To minimise as much as possible the systematic errors associated to the atmosphere evolution in time, its properties have to be continuously monitored. It is to this end that extensive atmospheric monitoring programs have been developed by different collaborations in astroparticle physics. Section~\ref{sec:astropart_exps} will list briefly the different experiments where the atmosphere is a part of the detector. In all cases, at some point, photons propagate into the atmosphere and they are affected by the medium before being detected. Section~\ref{sec:atmo_effects} will describe the different physics phenomena affecting photon propagation in the atmosphere in order to remove their effect in measurements. Then, in Section~\ref{sec:atmo_facilities}, the main instruments used to monitor the atmospheric properties or the atmosphere components will be presented. Astroparticle physics experiments, equipped with such infrastructures and located in unusual places, provide an opportunity to develop interdisciplinary activities, especially in atmospheric science and geophysics: this will be the purpose of Section~\ref{sec:interdisciplinary}.

\section{Astroparticle physics experiments and the atmosphere as part of the detector}
\label{sec:astropart_exps}
Astroparticle physics is a research field at the intersection of particle physics, astrophysics and cosmology. The term ''astro'' refers to the messengers from the universe and arriving on Earth. This area has the particularity to express the relationship increasingly close between the infinitely large (such as astrophysical objects in the universe) and the infinitely small (such as the study of the structure of matter). Origins of the field bring us back one century ago with the discovery in 1912 by Victor Hess (Nobel Prize in Physics in 1936,~\cite{Hess_overview}) of cosmic rays, opening at that time a new window for particle physics. Nowadays, astroparticle physics tries to answer three main issues: the role of high-energy phenomena in the universe, the composition of the universe -- only 5\% of the universe are known, the rest being composed of 26\% of dark matter and 69\% of dark energy whose origin and nature remain to be determined --, and the fundamental interactions at the ultimate energies.

The main messenger used to probe the universe is the photon, with a wavelength spreading from about $0.1~$nm (gamma-rays, X-rays) to a few kilometres (long-wavelength radio waves). Other messengers from universe can also be detected as cosmic rays or neutrinos: in some cases, photons produced by the interaction of these primaries with the atmosphere are recorded to evaluate indirectly the messenger properties. Whether it is direct or indirect messenger detection, photon propagation in the atmosphere is of principal interest in astroparticle physics. This section presents briefly the actual and future experiments using the atmosphere as part of their detector.

\subsection{Very-high energy gamma rays and ultra-high energy cosmic rays: a detection via extensive air showers}
\label{sec:airshowers}
The flux of ultra-high energy (UHE, $E \geqslant 10^{18}~$eV) cosmic rays and very-high energy (VHE, $E \geqslant 10^{11}~$eV) gamma rays is very low on Earth. To enlarge the detection area of these messengers from the universe, telescopes are directly installed on the ground and the Earth's atmosphere acts as the calorimeter of the detector. When cosmic rays or gamma rays enter the atmosphere, they induce extensive air showers composed of secondary particles. Among these particles, photons are emitted: their properties provide a direct way to probe the characteristics of the primaries. In the following, we describe the main experiments and their techniques employed to detect ultra-high energy cosmic rays or very-high energy gamma rays.  

\subsubsection{Ultra-high energy cosmic rays and fluorescence telescopes}
\label{sec:uhecr}
\begin{figure*}[!t]
\centering
\vspace{-0.7cm}
\resizebox{0.49\textwidth}{!}{%
\includegraphics{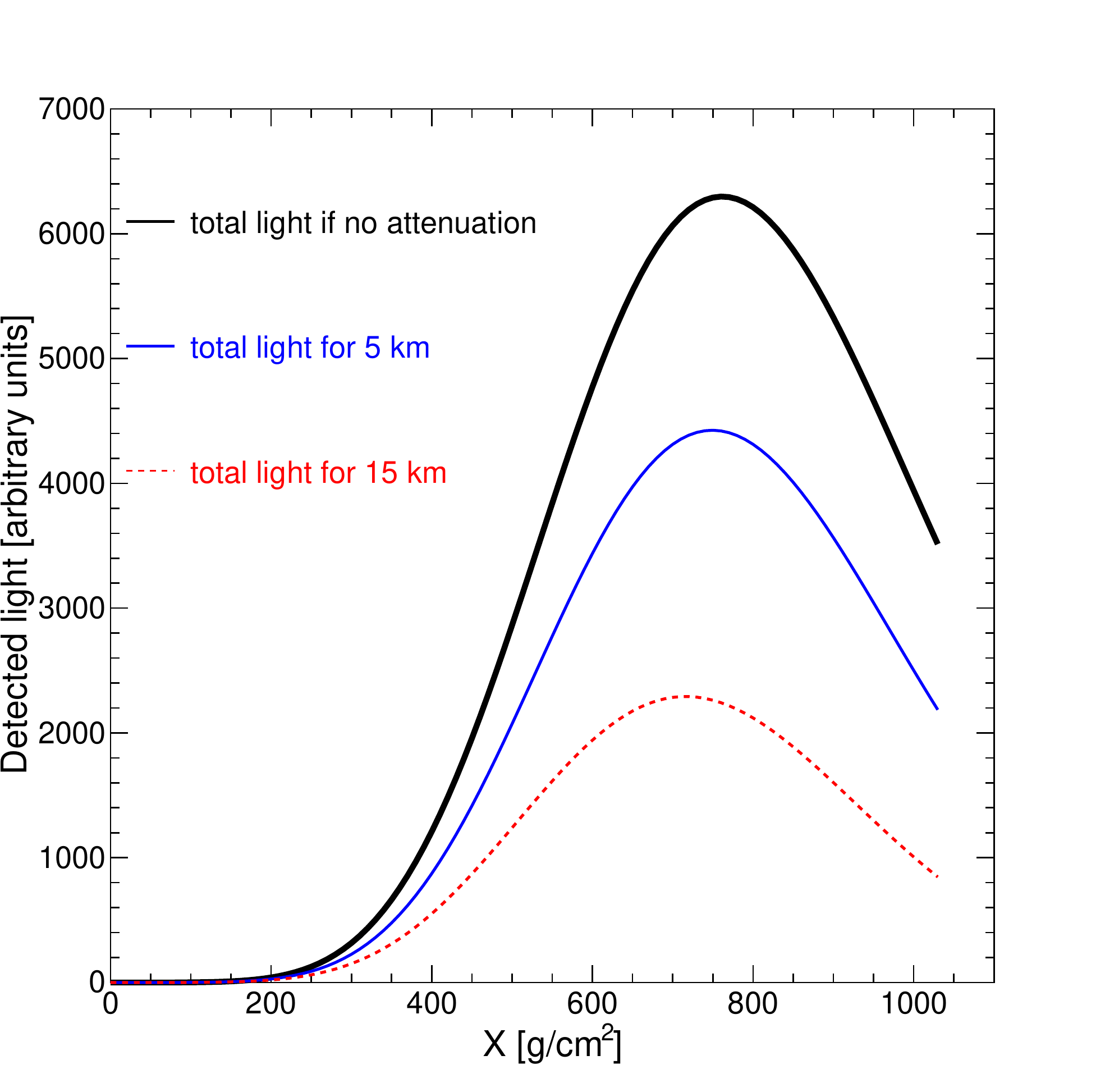}
}
\hfill{}
\resizebox{0.4\textwidth}{!}{%
\includegraphics{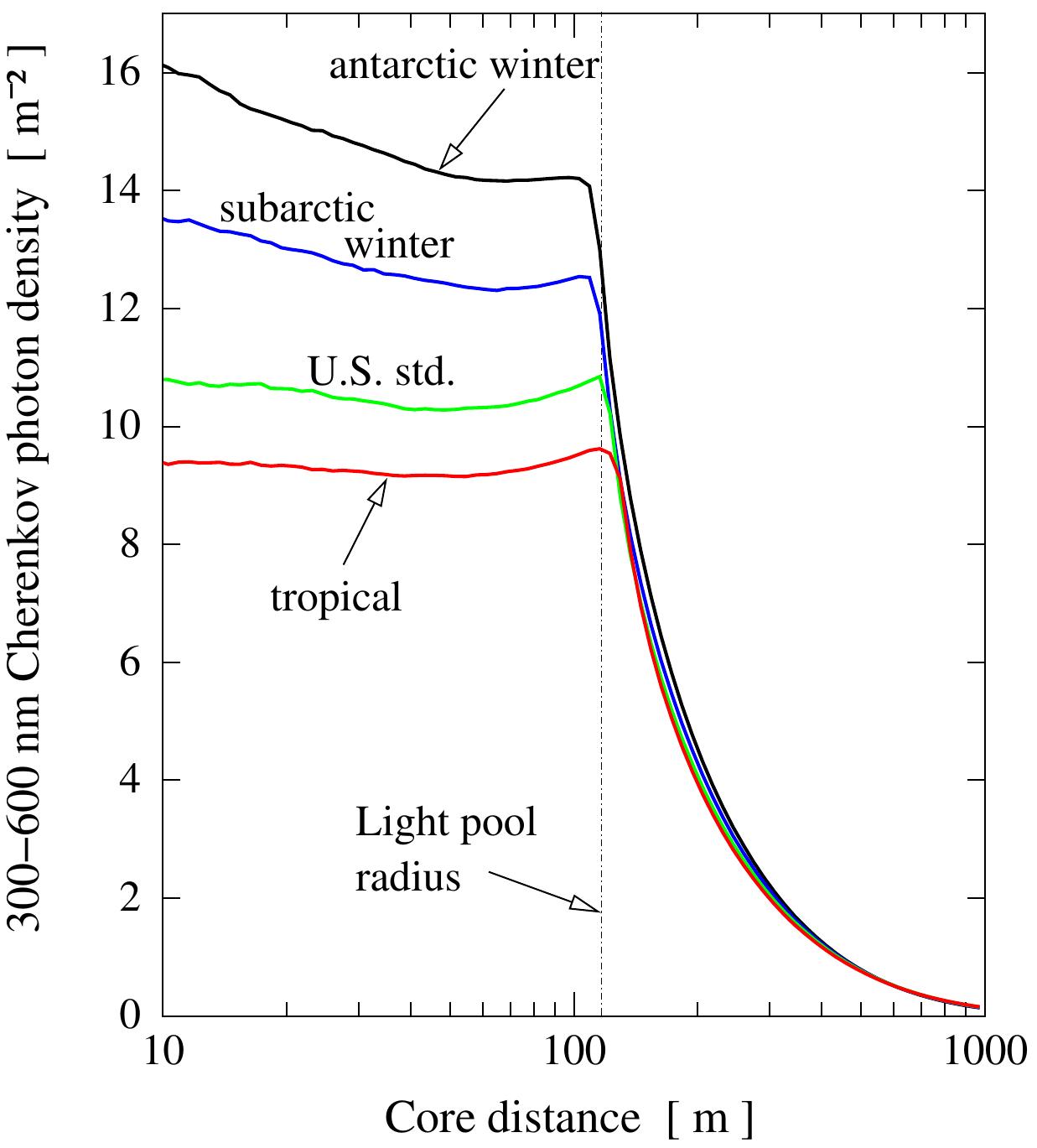}
}
\caption{(left) Reconstruction of the longitudinal profile for a vertical air shower induced by an UHECR at $5$ and $15~$km. The thick line represents the original longitudinal profile before any atmospheric attenuation. (right) Average lateral profiles of Cherenkov light density at ground level for different atmospheric profiles. Air showers are produced by a vertical 100 GeV gamma-ray and are observed at an observation level of $2\,200~$m above seal level (from~\cite{AtmoHEAD_Bernlohr}).}
\label{fig:astropart_atmo}
\end{figure*}

The cosmic ray energy spectrum observed on Earth extends from below $1~$GeV to beyond $10^{20}~$eV, more than eleven orders of magnitude. This energy spectrum drops rapidly with energy. For the so-called ultra-high energy cosmic rays corresponding to the right-hand limit of the spectrum, fundamental properties such as their origin, their chemical composition and their acceleration mechanisms are still a mystery (see~\cite{WalterEtAl,Watson,Berezinsky,Lemoine} for more details). At energies greater than $10^{18}~$eV, their flux is lower than one particle per century and per square kilometre. This makes these events only detectable indirectly through extensive air showers. Charged particles composing the air shower excite atmospheric nitrogen molecules, and these molecules then emit fluorescence light isotropically in the $300-420~$nm range~\cite{AIRFLY,ArquerosFY}. Detection of ultra-high energy cosmic rays using nitrogen fluorescence emission is a well established technique~\cite{sokolsky}, used in the past in the Fly's Eye~\cite{FlysEye} and HiRes~\cite{HiRes} experiments, currently at the Pierre Auger Observatory~\cite{pao,pao_fd} and Telescope Array~\cite{ta,ta_fd}, and in the future by the JEM-EUSO telescope \cite{jemeuso}. The energy and the geometry of extensive air showers can be calculated from information on the amount and the arrival time of recorded light signals at the fluorescence detectors (FD). After more than thirty years of development having led to a better understanding of this technique, the current ''hybrid'' observatories set their energy scale using fluorescence measurements~\cite{ICRC_Verzi,ICRC_Ikeda}. Also, the air-fluorescence technique allows the determination of the depth of maximum of the extensive air shower $X_{\rm max}$ in a direct way, providing an estimation of the UHECR composition~\cite{Auger_Xmax,TA_Xmax}. During the development of an extensive air shower, the production rate of fluorescence photons depends on the temperature, pressure and humidity in the air~\cite{BiancaFY,VazquezFY,DelphinePHIL}. For the greatest energies, fluorescence light from an air shower can be recorded at distances up to about $40~$km, traversing a large amount of atmosphere before reaching the detector: the atmospheric effects on the light propagation must hence be considered carefully (see Fig.~\ref{fig:astropart_atmo}(left)). To fulfil this task, extensive atmospheric monitoring programs are developed by these collaborations. In UHECR experiments, atmospheric data are not only used to reject periods with bad weather conditions but are directly applied in the reconstruction of extensive air showers at the observatory~\cite{AugerATMON,TaATMON}. Evaluating carefully atmospheric effects lead to a higher duty cycle for these experiments and improve the systematic errors in air shower reconstruction, two essential parameters in this research field where events are rare and require very large detectors.

\subsubsection{Very-high energy gamma rays and imaging atmospheric Cherenkov telescopes}
\label{sec:vhegr}
Very-high energy gamma rays are useful messengers to probe the populations of high-energy particles in distant regions or in our own galaxy. They are currently the easiest way to directly measure the most energetic phenomena in the universe such as supernova shock waves, pulsar nebulae, or active galaxy nuclei. Knowledge in this energy range was considerably increased during the last twenty five years, mainly achieved with ground-based Cherenkov telescopes~\cite{Hinton,Hillas,Bell,Aharonian,Gamma_overview}. Once a VHE gamma ray enters the atmosphere, secondary electrons, positrons and photons are generated. The propagation of electrons and positrons through the atmosphere at speeds greater than the speed of light produces a Cherenkov radiation beamed with respect to the shower axis. Then, this light is collected by imaging air Cherenkov telescopes (IACT) or air shower detectors -- the latter will not be discussed in the rest of this paper since the atmosphere is not a predominant effect on measurements. The main interests of this type of detector are a good rejection of the overwhelming cosmic ray background, the low energy threshold, and the angular and energy resolutions. This technique is currently used by the H.E.S.S.~\cite{HESS}, MAGIC~\cite{MAGIC} and VERITAS~\cite{VERITAS} collaborations and, in a near future, will be employed in the CTA project~\cite{CTA,CTA_concept} to cover the energy range from a few tens of GeV to a few hundreds of TeV. Cherenkov light is emitted through a cone where the cosine of its opening angle is equal to $1/(n\,\beta)$, where $n$ is the index of refraction and $\beta=v/c$ with $v$ being the velocity of the emitting particle. Also, the Cherenkov yield is directly linked to the refractive index of air. Thus, during the development of an electromagnetic air shower, a good knowledge of the atmospheric vertical profiles is needed~\cite{AtmoHEAD_Bernlohr}. Figure~\ref{fig:astropart_atmo}(right) shows the average Cherenkov light density at ground for typical gamma-ray showers, for different profiles of atmosphere. Once the Cherenkov emission is well evaluated, atmospheric extinction is another source of concern affecting directly the energy threshold and biases any flux measurement of astrophysical sources since misreconstructed energies shift the entire spectrum to lower energies. Whereas up to now data taking periods with bad atmospheric conditions are simply rejected in imaging air Cherenkov experiments, there are some studies evaluating the feasibility to correct data recorded also during non-optimal atmospheric conditions \cite{AtmoHEAD_Rulten,AtmoHEAD_Nolan}. Attempts for such a result applied to the next generation of imaging air Cherenkov experiments CTA are presented in~\cite{AtmoHEAD_Rulten}: after having recorded a softer energy spectrum of an astrophysical source, measurements of atmospheric attenuation \emph{in situ} would permit to obtain a correction factor to apply to the spectrum in order to come back to the original one. Whereas this practice is now common in experiments detecting ultra-high energy cosmic rays (see Sect.~\ref{sec:uhecr}), reconstruction of air showers even in unclear atmospheric conditions is not yet widespread in ground-based gamma astronomy.

\subsection{Astronomical all-sky surveys and the ground-based photometric measurements}
\label{sec:astronomy}
The twentieth century has been the emergence of the standard model of cosmology to describe the evolution of the universe since the Big Bang. Through multiple experimental probes such as the large scale distribution of galaxies, the study of supernovae or the cosmic microwave background, this model also called $\Lambda$CDM (Cold Dark Matter) is imposed over the years. Great advances in our understanding of the universe came from large scale sky surveys in many wavebands, such as SDSS~\cite{SDSS}, SNLS~\cite{SNLS}, 2MASS~\cite{2MASS} or FIRST~\cite{FIRST}. Despite its many successes in explaining measurements until today, only 5\% of the composition of the universe are known, the rest consisting of 26\% of dark matter acting only through the gravitational force and 69\% of dark energy whose origin remains to be determined and is described by the cosmological constant $\Lambda$. The next generation of instruments in the field aims to better understand the nature and the origin of dark matter and dark energy, using mainly the following cosmic probes: the weak lensing cosmic shear of galaxies, the baryon acoustic oscillations in the power spectrum of galaxy distribution, and the relationship between redshifts and distances for type Ia supernovae~\cite{LSST}. The simultaneous study of these probes on the same data set can check the consistency of different cosmological models describing the universe. The Euclid satellite~\cite{Euclid} -- from space, i.e.\ without atmosphere -- and the ground-based large-area surveys such as the Dark Energy Survey~\cite{DES}, Pan-STARRS~\cite{Pan-STARRS} and the Large Synoptic Survey Telescope~\cite{LSST_web} hold the first places to fill this role.

A ground-based telescope with a broad-band detector -- from $300~$nm to $1100~$nm -- records the integral of the source specific flux density at the top of the atmosphere, weighted by the response function depending on the effects of the atmosphere and the instrumental optics. The science goals are achieved thanks to multiple images of the sky recorded over the course of many years, in less than ideal measurement conditions. In this case, it is challenging to obtain a calibration of broad-band photometry stable in time and uniform over the sky to precisions of 1\%~\cite{Ivezic_1}. This goal can be reached only if an extensive work is undertaken to monitor continuously the optical transmittance from the top of the atmosphere to the input pupil of the telescope~\cite{Stubbs_1,Burke_1,Burke_2} and the instrumental system response from the input pupil to the detector~\cite{Stubbs_2}. 

\section{Atmospheric effects on light production and / or its propagation in the atmosphere}
\label{sec:atmo_effects}
\begin{figure*}[!t]
\centering
\resizebox{0.49\textwidth}{!}{%
\includegraphics{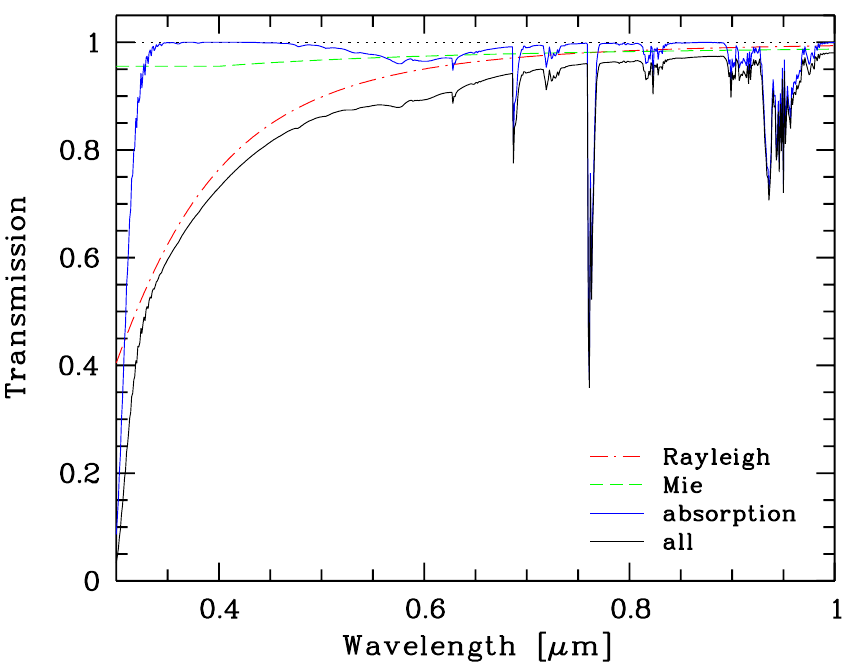}
}
\hfill{}
\resizebox{0.49\textwidth}{!}{%
\includegraphics{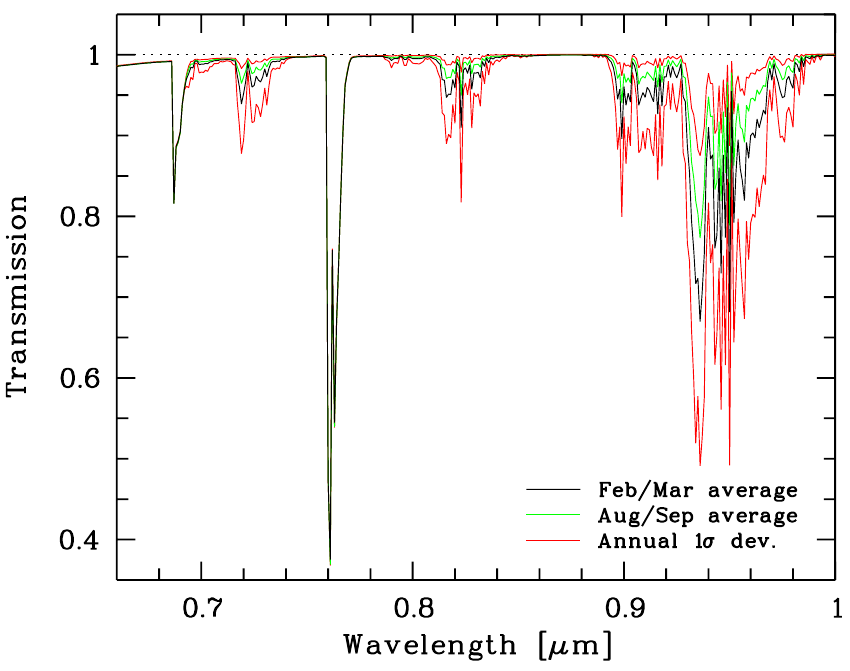}
}
\caption{(left) Annual mean zenith transmission curve at Cerro Paranal, Chile ($24^\circ$ 38' S, $70^\circ$ 24' W, $2635~$m ASL). Absorption contribution is mainly due to molecular oxygen, water vapour and ozone. (right) Variation of molecular absorption at Cerro Paranal. The best transmission curve is found in August/September, i.e.\ end of Austral winter, and the lowest one is in February/March, i.e.\ end of Austral summer (from~\cite{Noll}).}
\label{fig:transmission_spectrum}
\end{figure*}

Although the atmosphere is globally opaque to electromagnetic radiation, two wavelength ranges permit the detection of photons: from about $300~$nm to a few tens of micrometre (ultraviolet / UV, visible and near-infrared / IR, where the $300-1100~$nm is designed as ''optical'' in the rest of this paper), and from a few centimetres to about $10~$m (centimetric to decametric radio waves). Whereas the radio wavelength domain presents a very clean atmospheric transmission, this is not exactly the case in UV, visible or near-IR where some distortions are present in the spectrum of atmospheric transmission (see Fig.~\ref{fig:transmission_spectrum} (left)). If the atmosphere is used as a giant calorimeter, it has been mentioned in Sect.~\ref{sec:airshowers} that the molecular component affects also the yield of fluorescence and Cherenkov lights. Cherenkov and fluorescence light production at a given wavelength $\lambda$ depends on the atmospheric variables pressure $P$, temperature $T$ and vapour pressure $e$. Whereas the Cherenkov light yield can be directly calculated from the refractive index of the atmosphere $n(\lambda,P,T)$, the weather dependence of the fluorescence production is much more complicated to determine. Among the effects being difficult to measure experimentally, we can cite the collisional quenching of fluorescence emission. In this phenomenon, the radiative transitions of excited nitrogen molecules are suppressed by molecular collisions. Also, water vapour contributes to collisional quenching leading to an additional dependence on the atmosphere humidity for the fluorescence yield~\cite{BiancaUHECR2012}. However, for all phenomena affecting the light production, a simple knowledge of the vertical profiles of temperature, pressure or vapour pressure is needed to well estimate the corresponding yields.

Attenuation of light from the production point to the detector can be expressed as a transmission coefficient $\Gamma$, or optical transmittance, giving the fraction of incident light at a specified wavelength $\lambda$ along the path length $x$. If $\tau$ is the optical depth, $\sigma$ the cross section and $N(x)$ the density of the component along the line of sight $x$, then $\Gamma$ is estimated using the Beer-Lambert law
\begin{equation}
\begin{split}
\Gamma(x,\lambda) &= (1 + H.O.) \sum_i \Gamma_{i}(x,\lambda)\\
& = (1 + H.O.) \exp\,\left[-\prod_i \tau_i(x,\lambda)\right]\\
& = (1 + H.O.) \exp\,\left[-\prod_i \int_0^x \sigma_i(\lambda)\,N_i(x) {\rm d}x \right],
\end{split}
\label{eq:beer}
\end{equation}
where $\{\Gamma_i$, $\tau_i$, $\sigma_i$, $N_i(x)\}$ represent the different contributions to the light attenuation in the atmosphere. They can be grouped into four categories: the molecular absorption, the molecular scattering, the aerosol scattering and the cloud extinction. The optical depth expresses the quantity of light removed from a beam by \emph{absorption} -- the radiant energy is transformed into other wavelengths or other forms of energy -- or \emph{scattering} -- the energy received is re-radiated at the same wavelength usually with different intensities in different directions -- during its path through a medium. Light is not only scattered out of the field of view of the detector and can be also scattered into it: $H.O.$ represents this higher-order correction taking into account for single and multiple scattering into the field of view of the detector (see Sect.~\ref{sec:mult_scattering}). The rest of this section consists in describing these four categories stressing different dependences on the atmosphere thickness $x$ and on the incident wavelength $\lambda$, and leading finally to the wavelength-dependent global optical transmission spectrum of atmosphere.

\subsection{Molecular absorption -- wavelength dependent}
\label{sec:molecular_absorption}
The air is a medium with a mass composed of $78\%$ dinitrogen ${\rm N}_{\rm 2}$ and $21\%$ dioxygen ${\rm O}_{\rm 2}$ (then, traces of argon~Ar, neon~Ne, helium~He, dihydrogen~${\rm H}_{\rm 2}$ and xenon~Xe). They correspond to the ''permanent gases'' composing the atmosphere. The resulting molecular mass for an ''air molecule'' is $M_{\rm 0} = 28.97~$g/mol for standard temperature and pressure at sea level. To take into account humidity effects, air must be added a factor corresponding to the water vapour. The final molecular mass with respect to the altitude above sea level (ASL) $h$ is the sum of the two components, weighted by their volume fractions
\begin{equation}
M_{\rm m}(h) = \alpha_{\rm dry}(h)\, M_{\rm 0} + \alpha_{\rm w}(h)\, M_{\rm w},
\end{equation}
where the molecular masses for dry air and water vapour are $28.97~$g/mol and $44.01~$g/mol, respectively. The atmosphere is not only composed of these permanent gases and water vapour H$_2$O, other ''variable gases'' are also present in a very small quantity in the atmosphere. Among them, the main gases are carbon oxides CO and CO$_2$, methane CH$_4$, ozone O$_3$, nitrogen oxides NO and NO$_2$, or sulfur dioxides SO$_2$. We have to consider also the Volatile Organic Compounds (VOCs) which are a class of organic compounds, most of them being hydrocarbons. Presently, theoretical models are not generally available to provide absorption cross sections for each species for different temperature and pressure values. The cross sections of the chemical species are measured in laboratory at fixed temperature and pressure, then empirical models are used to interpolate them to intermediate values of temperature and pressure. Nowadays, the largest molecular spectroscopic database is the HITRAN (HIgh resolution TRANsmission) database serving as input for radiative transfer calculation codes~\cite{HITRAN_web,HITRAN_1,HITRAN_2}. After a brief listing of the main gases present in the atmosphere, an estimation of their contribution to the optical transmission spectrum of atmosphere is done in the following.

Light absorption in the optical wavelength range is dominated by three atmospheric gases: molecular oxygen ${\rm O}_{\rm 2}$, water vapour H$_2$O and ozone O$_3$. Absorption by {\bf molecular oxygen O$_2$} presents three narrow absorption bands in the transmission spectrum: the Fraunhofer A band at $760~$nm, the Fraunhofer B band at $690~$nm and the $\gamma$ band at $630~$nm. At shorter wavelengths, the atmosphere is strongly opaque due to the bands of the Schumann-Runge system between $175~$nm and $200~$nm, and the weak Herzberg dissociation continuum extending from $175~$nm to $260~$nm. Since molecular oxygen is a well-mixed gas in the atmosphere, intensity of these bands depends only on the atmospheric density and it is axisymmetric with respect to the zenith. {\bf Ozone O$_3$} is a triatomic molecule far less stable than molecular oxygen. About 97\% of the total ozone column is found in the stratosphere, the so-called {ozone layer}, where it is produced naturally via the photo-dissociation of the molecular oxygen by UV radiation, followed by a recombination of oxygen molecules and oxygen atoms. A small fraction of ozone is also located at the Earth's surface, i.e.\ in the troposphere, produced in the smog formed in large cities where the presence of sunlight generates photo-chemical reactions. The opacity of ozone is responsible of the total loss of atmospheric transmission below $300~$nm with the Hartley and Huggins bands. Between $500~$nm and $700~$nm, the broad Chappuis bands attenuate light of a few per cent. Since ozone is mainly located in the stratosphere, its temporal and spatial variability is low. This characteristic is not at all the case for {\bf water vapour H$_2$O} in the atmosphere, a constituent not well-mixed in the air. Water vapour is mainly found in the lowest part of the atmosphere. However, a ground-level value of relative humidity is not a right estimate of the total column height. It absorbs electromagnetic radiation in the optical wavelength range through different bands, where the prominent ones are at $720~$nm, $820~$nm and $940~$nm. Due to the high variability in time and space of the water vapour component, intensity of these bands varies and a continuous monitoring is advised (see Sect.~\ref{sec:exp_molecular}). Figure~\ref{fig:transmission_spectrum}(right) depicts this phenomenon: whereas absorption bands related to molecular oxygen are similar in summer and winter, amplitude of bands due to water vapour varies.

Additional trace gases are also present in the atmosphere and could absorb a part of the electromagnetic radiation since their absorption cross section is not negligible in optical wavelength range. These gases are not present in a same quantity in every locations and require a specific measurement program to monitor them. These atmospheric trace gases come from both natural sources and human activities. Examples of natural sources include wind picking up dust and soot from the Earth's surface and carrying it aloft, volcanoes belching tons of ash and dust into our atmosphere, or forest fires producing vast quantities of drifting smoke. As human-induced sources, one cites usually transportation, fuel combustion or industrial processes. These different gases had previously constant concentration in the atmosphere, since their origin was exclusively natural. But, nowadays, human activities increase the concentration of these gases called air pollutants. Among trace gases present in the atmosphere but not affecting the transmission spectrum in the optical wavelength range, we can list carbon oxides CO (absorbing in near-infrared) and CO$_2$ (absorbing in near-ultraviolet), methane CH$_4$ (absorbing in near-infrared) and nitric acid HNO$_3$ (absorbing in near-ultraviolet). On the contrary, other trace gases affect strongly the transmission spectrum if they are found in enough quantity in the atmosphere. {\bf Sulfur dioxide SO$_2$} is a colourless gas coming primarily from the burning of sulfur-containing fossil fuels (such as coil or oil). It can enter the atmosphere naturally during volcanic eruptions and as sulfate particles from ocean spray. Its corresponding absorption band in the ultraviolet is spread from $400~$nm to $700~$nm, depicting very sharp rotational structures visible only at high spectral resolution~\cite{SO2}. {\bf Nitrogen dioxide NO$_2$} is a gas formed when some of the nitrogen in the air reacts with oxygen during the high-temperature combustion of fuel. Although it is produced naturally, its concentration in urban environments is 10 to 100 times greater than in non-urban regions. NO$_2$ absorption covers a large part of the entire optical spectrum from near-ultraviolet to near-infrared, with a peak reached about $400~$nm~\cite{NO2}. In moist air, nitrogen dioxide reacts with water vapour to form corrosive nitric acid HNO$_3$, a substance that adds to the problem of acid rain. Moreover, nitrogen dioxide is highly reactive and plays a key role in producing ozone and other ingredients of photo-chemical smog. To a lesser extent, the {\bf nitrate radical NO$_3$} could also affect the transmission of electromagnetic radiation in visible via the strongest absorption feature at $660~$nm~\cite{NO3}. In the troposphere, the main nighttime oxidant is the nitrate radical NO$_3$ formed by the relatively low oxidation of NO$_2$ by O$_3$~\cite{NO3_radical}.

\subsection{Molecular scattering -- wavelength dependent}
\label{sec:molecular_scattering}
The molecular component of the atmosphere is described by the height-dependent profiles of its state variables pressure $P(h)$ and temperature $T(h)$, where $h$ corresponds to the altitude above sea level. These vertical profiles can be provided with balloon-borne radiosonde flights or with atmospheric models from numerical weather predictions (see Sect.~\ref{sec:exp_molecular}). As a first approximation, the air density in molecules per m$^3$ as a function of height above sea level $N_{\rm m}(h)$ can be parameterised as
\begin{equation}
\begin{split}
N_{\rm m}(h) &= \frac{N_{\rm A}}{R}\,\frac{P(h)}{T(h)} = N_{\rm m|s}\, \frac{T_{\rm m|s}}{P_{\rm m|s}}\, \frac{P(h)}{T(h)}\\
& = \frac{N_{\rm A}}{R}\,\frac{P(0)}{T(0)}\, \exp\left(-\frac{h}{H_{\rm m}} \right),
\end{split}
\label{eq:transmission_ray_0}
\end{equation}
where $R$ is the universal gas constant, $N_{\rm A}$ the Avogadro constant, and $H_{\rm m}$ the scale height for molecular component ($H_{\rm m} \simeq 8~$km). Standard air is defined by the temperature $T_{\rm m|s}= 288.15 ~$K, the pressure $P_{\rm m|s}=1\,013.25\times10^2~$Pa and the molecular density $N_{\rm m|s}=2.54743\times 10^{25}$ m$^{-3}$. Water vapour H$_2$O is not a well-mixed constituent of the molecular component and does not follow the same dependence on the altitude~\cite{Parameswaran}. Of course, almost all the physical quantities given in Eq.~\eqref{eq:transmission_ray_0} vary with time and require a continuous monitoring. In the same way, the refractive index of dry air $n_{\rm m}$ for incident wavelengths greater than $230~$nm in function of the altitude $h$ is given by~\cite{Birch}
\begin{equation}
\begin{split}
&n_{\rm m}(\lambda, h) - 1 = [n_{\rm m|s}(\lambda) - 1]\times\frac{P(h) / {\rm Pa}}{96\,095.43}\\
& \times \frac{1 + 10^{-8}\left[0.613 - 0.009\,98\, T(h)/{^\circ {\rm C}} \right]\, [P(h) / {\rm Pa}]}{1 + 0.003\,661 \, T(h)/{^\circ {\rm C}}},
\end{split}
\label{eq:transmission_ray_1}
\end{equation}
where the refractive index for an atmosphere in the standard temperature and pressure conditions is defined by the formula~\cite{PeckReeder}
\begin{equation}
\begin{split}
n_{\rm m|s}(\lambda)& = 1 + \frac{0.057\,918\,17}{238.0185-(1\,\upmu{\rm m}/\lambda)^2}\\
&+ \frac{0.001\,679\,09}{57.362-(1\,\upmu{\rm m}/\lambda)^2}.
\label{eq:transmission_ray_2}
\end{split}
\end{equation}

Molecular scattering from near-UV to near-IR wavelengths can be approximated using uniquely the elastic Rayleigh scattering process, since it dominates inelastic Raman scattering (by around three orders of magnitude) or absorption~\cite{Burris}. Rayleigh theory can be applied if scattering particles are much smaller than the light wavelength. The Rayleigh scattering cross section per air molecule is given analytically by the following formula~\cite{Bucholtz}
\begin{equation}
\begin{split}
\sigma_{\rm R}(\lambda) &= \frac{24\pi^3}{\lambda^4 {N_{\rm m|s}}^2}\left[\frac{{n_{\rm m|s}}^2(\lambda)-1} {{n_{\rm m|s}}^2(\lambda)+2}\right]^2 \frac{6+3\,\delta_{\rm n}(\lambda)} {6-7\,\delta_{\rm n}(\lambda)}\\
& = A\,\lambda^{-\left(B + C\,\lambda + D/\lambda\right)},
\end{split}
\label{eq:transmission_ray_3}
\end{equation}
where $n_{\rm m|s}(\lambda)$ is the refractive index for standard air and $\delta_{\rm n}(\lambda)$ the depolarisation factor taking into account the anisotropy of the air molecules. Values for $\{A$, $B$, $C$, $D\}$ can be found in~\cite{Bucholtz}. $\delta_{\rm n}(\lambda)$ is determined by the asymmetry of N$_2$ and O$_2$ molecules, and equal to zero for point-like scattering centres~\cite{Bucholtz}. The depolarisation factor varies by approximately $60\%$ from the near-IR ($\delta_{\rm n}\simeq0.045$) to the UV domain ($\delta_{\rm n}\simeq0.028$), introducing a corresponding variation with wavelength of around $3\%$ for the Rayleigh scattering cross section. The value chosen leads to a shift for the wavelength dependence of the molecular scattering from the well-known $\lambda^{-4}$ behaviour to an effective value of $\lambda^{-X}$. Equations~\eqref{eq:transmission_ray_0}--\eqref{eq:transmission_ray_3} are given in the case of a dry air. However, A. Bucholtz shown in~\cite{Bucholtz} that the quantity ${({n_{\rm m|s}}^2-1)}$ / ${({n_{\rm m|s}}^2+2)}$ varies of less than 1\% for a typical water vapour density of $7~$g/cm$^2$. Thus, calculations of the Rayleigh scattering cross section without taking into account water vapour content in the atmosphere does not lead to a wrong approximation. Finally, the molecular optical depth integrated from the sea level up to an altitude $h$ and observed through a zenith angle $\theta$ is obtained with 
\begin{equation}
\begin{split}
\tau_{\rm m}(h,\theta,\lambda) &= (1/\cos\theta)\int_0^h \sigma_{\rm R}(\lambda)\, N_{\rm m}(h') {\rm d}h'\\
& = (1/\cos\theta)\int_0^h \alpha_{\rm m}(\lambda,h') {\rm d}h',
\end{split}
\label{eq:transmission_ray_4}
\end{equation}
i.e.\ an attenuation axisymmetric about the zenith ($\theta=0^\circ$) with a time dependence driven only by pressure and temperature variations. $\alpha_{\rm m}(\lambda,h)$ is the so-called molecular extinction coefficient.

Due to the limited field of view of detectors, a non-negligible fraction of photons are detected after one or several scatterings and scattering properties of the atmosphere need to be well estimated. A scattering phase function is used to describe the angular distribution of scattered photons. It is typically written as a normalised probability density function expressed in units of probability per unit of solid angle. When integrated over a given solid angle $\Omega$, a scattering phase function gives the probability of a photon being scattered with a direction that is within this solid angle range. Since scattering is always uniform in azimuthal angle $\phi$ for both aerosols and molecules, the scattering phase function is always written simply as a function of polar scattering angle $\zeta$. Molecules are governed by Rayleigh scattering that can be derived analytically via the approximation that the electromagnetic field of incident light is constant across the small size of the particle. The molecular phase function, symmetric in the forward-backward direction, is proportional to the $(1 + \cos^2\zeta)$ factor. Due to the anisotropy of the N$_2$ and O$_2$ molecules, a small correction factor $\delta$ is included and is equal to about one per cent in the case of air~\cite{Bucholtz},
\begin{equation}
P_{\rm m}(\zeta|\delta) = \frac{3}{16\pi(1+2\,\delta)}\left[(1+3\,\delta) + (1-\delta)\cos^2\zeta \right],
\label{eq:angular_dependence_1}
\end{equation}
where $\zeta$ is the polar scattering angle, $P_{\rm{m}}$ the probability per unit solid angle, and the depolarisation factor $\delta_{\rm n}$ is part of the new parameter $\delta = \delta_{\rm n} / (2 - \delta_{\rm n})$.

\subsection{Aerosol scattering -- wavelength dependent}
\label{sec:aerosol_scattering}
Although atmosphere is mainly composed of molecules, a small fraction of larger particles such as dust, smoke, sea salt or droplets are in suspension. The aerosol component is defined with respect to the ground level $h_{\rm gl}$ instead of the sea level. These particles are called {\it aerosols} and their typical size varies from $1~$nm to about $100~\upmu$m. Atmospheric particle size distribution is generally a continuous multi-modal function spanning up to ten or more decades of concentration, usually expressed in terms of superimposed lognormal distributions~\cite{Whitby1978}. Each distribution represents a mode having a chemically distinct composition due to a specific source: the ''nucleation'' mode for aerosol sizes between $1~$nm to about $0.1~\upmu$m, the ''accumulation'' mode between $0.1~\upmu$m and $1.0~\upmu$m, and the ''coarse'' mode for sizes greater than $1.0~\upmu$m. Due to the hygroscopic nature of atmospheric aerosols, relative humidity affects their size: an increase in humidity will grow the aerosol size, especially for relative humidity values greater than 50\%. Whereas fine modes originate mainly from condensation sources and atmospheric gas-to-particle conversion processes, mechanical processes as wind driven soil erosion or seawater bubble-bursting produce the coarse mode. The latter mode has a considerably shorter atmospheric residence time than the sub-micrometre aerosol fraction. Concerning their effect on radiative processes, the accumulation mode is the most important since it represents the size range in which light scattering is the most efficient. However, in regions impacted by high levels of coarse mode aerosols such as deserts -- soil dust -- or the marine boundary layer -- sea salt --, coarse mode aerosols represent most of the total mass and affect in a non-negligible way the solar radiation scattering. On the contrary, the nucleation mode represents just a small part of the total aerosol mass and is inefficient on light scattering processes.

Since the size of these particles is no longer small with respect to the incident wavelength, the analytical Rayleigh formulae cannot be applied in this case. The Mie scattering that they produce is much more complex~\cite{Mie}: for instance, it depends on particle composition, particle shape and on the particle size distribution. Also, aerosol populations, unlike the molecular component, change quite rapidly in time, depending on the wind and weather conditions. Two main physical quantities have to be known to evaluate the effect of the aerosols on photon propagation in the atmosphere: the height-dependence of aerosol optical depth $\tau_{\rm a}(h,\theta,\lambda)$ and the aerosol scattering phase function $P_{\rm a}(\zeta)$. Dust, in particular soot, also absorbs light: the fraction of absorbed light is given by the so-called {single scattering albedo}, equal to the ratio of the scattering cross section over the extinction (i.e.\ total) cross section. It evaluates the probability that a photon is scattered, rather than absorbed, during an interaction with an aerosol particle. However, this probability is usually close to one and aerosols are approximated to scattering centres only. 

The scattering cross section by these particles is less easily described as the electromagnetic field of incident light can no longer be approximated as constant over the volume of the particle~\cite{Hulst,Bohren}. The $\lambda^{-4}$ dependence of the total scattered intensity found by Rayleigh is no longer valid for the Mie solution, giving rise to a cross section $\sigma_{\rm a}(\lambda)$ scaling less strongly with wavelength than does molecular scattering ($\lambda^{-1}$ to $\lambda^{-2}$). Mie scattering theory offers a solution in the form of an infinite series for the scattering of non-absorbing spherical objects of any size. The number of terms required in this infinite series to calculate the scattering cross section is given in~\cite{Wiscombe}. Therefore height-dependent profiles of the vertical aerosol optical depth $\tau_{\rm a}(h,\lambda)$ are usually measured to evaluate aerosol population in the atmosphere. In this way, an indirect measurement of the aerosol size distribution and of the aerosol concentration is obtained. Aerosols are mainly found is the lowest part of the atmosphere, the so-called {planetary boundary layer}, whose the behaviour is directly influenced by friction and by heat fluxes with the Earth's surface. The planetary boundary layer has a thickness quite variable in space and time, usually around $1~$km but it can vary from $100~$m to about $3~$km. A characteristic shape of the vertical profile of $\tau_{\rm a}(h,\lambda)$ is the following: a more or less linear increase at the beginning, then a flattening since the aerosol concentration decreases quickly with the altitude. To evaluate the aerosol extinction for a given incident wavelength $\lambda$ and observed through a zenith angle $\theta$, a common parameterisation used is a power law
\begin{equation}
\begin{split}
\tau_{\rm a}(h,\theta,\lambda) &=  (1/\cos\theta)\int_{h_{\rm gl}}^h \alpha_{\rm a}(\lambda,h') {\rm d}h'\\
& = (1/\cos\theta)\, \tau_{\rm a}(h,\lambda_0)\, \left( \frac{\lambda_0}{\lambda}\right)^\gamma,
\end{split}
\label{eq:transmission_mie_1}
\end{equation}
where $\lambda_0$ is the wavelength used for the measurement of $\tau_{\rm a}(h,\lambda_0)$, $\alpha_{\rm a}(\lambda,h)$ is the so-called aerosol extinction coefficient and $\gamma$ is known as the Angstr\"om coefficient~\cite{Angstrom}. This exponent depends on the size distribution of aerosols~\cite{Angstrom_1,Angstrom_2}. When the aerosol particle size approaches the size of air molecules, $\gamma$ should tend to $4$ (mainly dominated by {accumulation-mode} aerosols), and for very large particles, typically larger than $1~\upmu$m, it should approach zero (dominated by {coarse-mode} aerosols). Usually, a $\gamma \simeq 0$ is characteristic of a desert environment and the aerosol optical depth is more or less independent of the incident wavelength. Thus, the exponent $\gamma$ is an indirect measurement of the aerosol size.

\begin{figure*}[!t]
\centering
\vspace{-0.7cm}
\resizebox{0.49\textwidth}{!}{%
\includegraphics{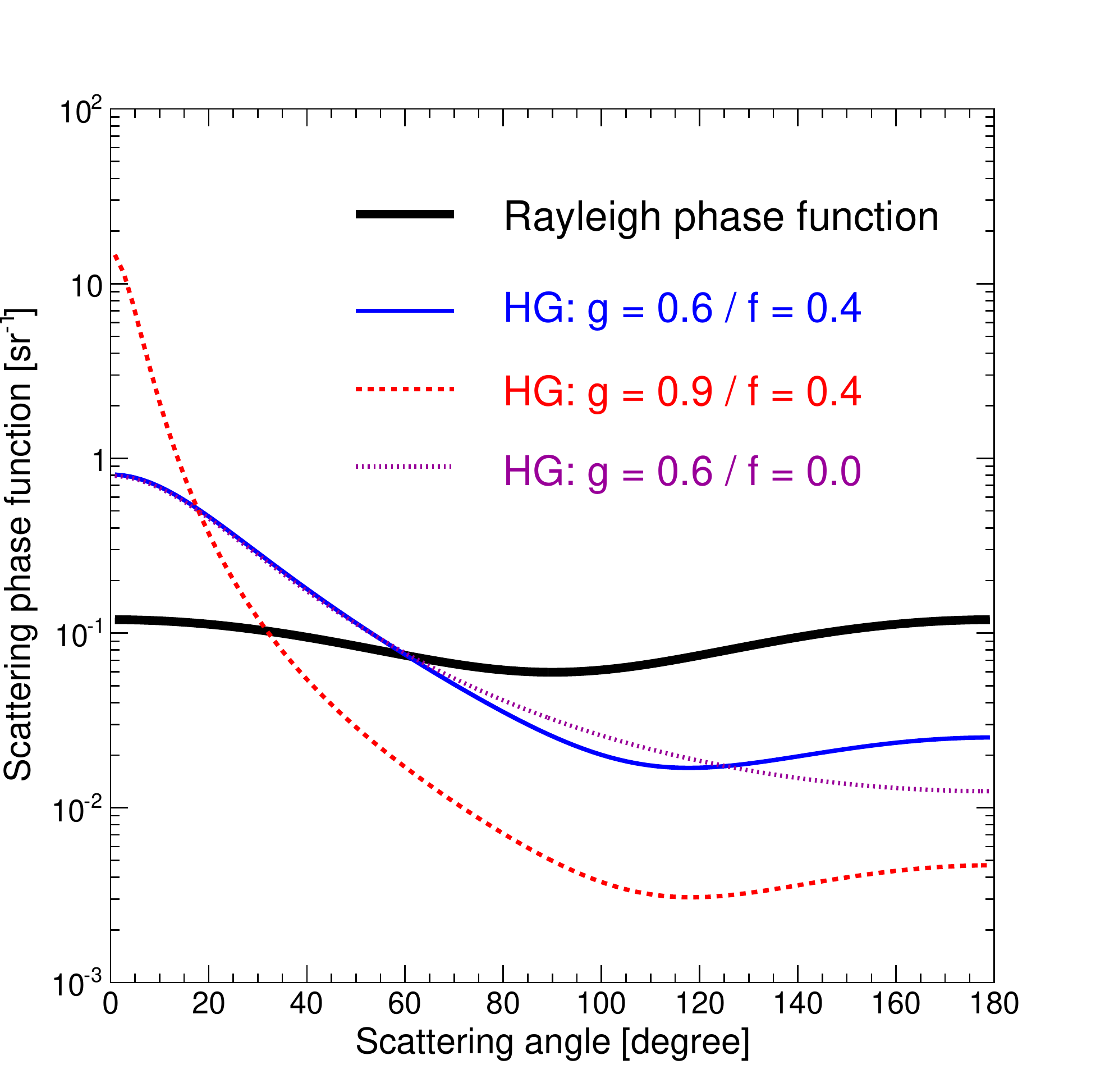}
}
\hfill{}
\resizebox{0.49\textwidth}{!}{%
\includegraphics{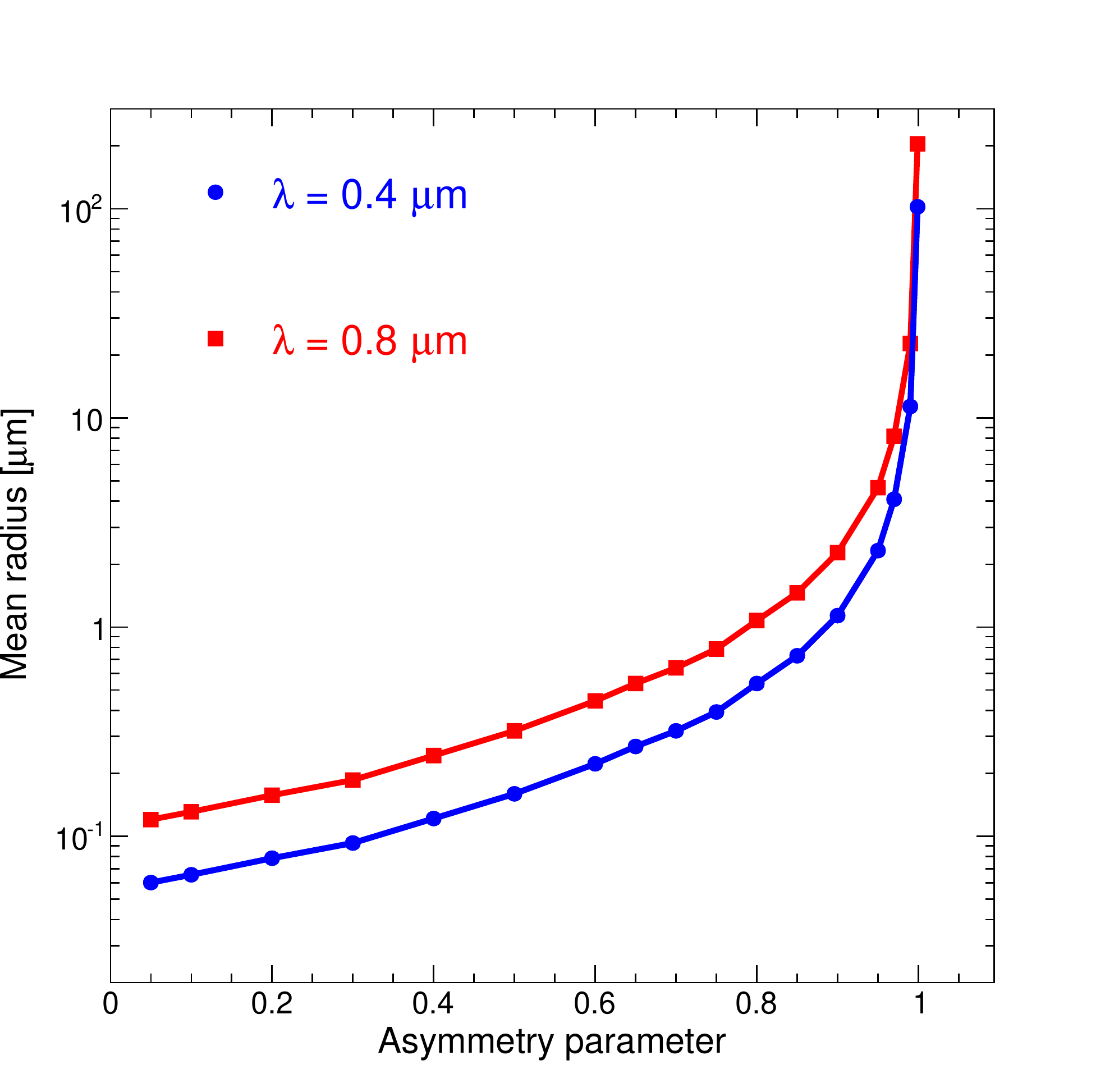}
}
\caption{(left) Henyey-Greenstein functions representing the scattering phase function for different asymmetry parameters $g$ and backward factors $f$. The Rayleigh scattering phase function, proportional to $(1+\cos^2\zeta)$, is also plotted. (right) Relation between the asymmetry parameter of the HG function and the mean radius of the particle size distribution. The equivalence is plotted for two different incident wavelengths: $0.4~\upmu$m (blue circles) and $0.8~\upmu$m (red squares) (from~\cite{Ramsauer_AO}).}
\label{fig:HenyeyGreenstein}
\end{figure*}

A much more direct estimation of the aerosol size is provided by the aerosol scattering phase function. It is well-known that the shape of the scattering phase function is dependent on the aerosol size. However, due to the Mie formalism, it is difficult to get a basic relationship between the two quantities. A new approach, the so-called Ramsauer approach, avoids this difficulty. The Ramsauer effect was discovered in 1921 and it is a model known in atomic and nuclear physics~\cite{Ramsauer}. Its main advantage is its intuitive understanding of an incident particle scattering over a sphere. Originally applied to electron scattering over atoms~\cite{Ramsauer_atoms} or neutron scattering over nuclei~\cite{Ramsauer_nuclei}, it can be also used for light scattering on non-absorbing spherical particles~\cite{Ramsauer_PhysScr,Ramsauer_AO}. This approach highlights the key role of the forward-scattering peak since the full-width at half maximum of this peak is proportional to the inverse of the aerosol size. Also, the amount of light scattered in the forward direction ($\zeta=0^\circ$) is proportional to the target area, i.e.\ square of the sphere radius. In other words, at constant wavelength, a larger aerosol scatters more light in the forward direction. One of the most popular scattering phase functions is the Henyey-Greenstein (HG) function~\cite{HenyeyGreenstein}. Henyey and Greenstein first introduced this function in 1941 to describe scattering processes in galaxies. This is a parameterisation usually used to reproduce scattering on objects large with respect to the incident wavelength, valid for various object types and different media~\cite{HG_astro,HG_meteo,HG_bio}. If the backscattering can not be neglected, the HG function becomes a ``Double HG'' and is given by
\begin{equation}
\begin{split}
&P_{\rm a}(\zeta|g,f) = \frac{1-g^2}{4\pi}\\
& \times \left[\frac{1}{(1+g^2-2g\cos{\zeta})^\frac{3}{2}} +f\left(\frac{3\cos^2{\zeta}-1}{2(1+g^2)^\frac{3}{2}}\right)\right],
\end{split}
\label{eq:APF}
\end{equation}
where $g$ is the asymmetry parameter given by $\left<\cos\zeta\right>$ and $f$ the backward scattering correction parameter. $g$ and $f$ vary in the intervals $[-1,1]$ and $[0 ,1]$, respectively. The parameter $f$ is an extra parameter acting as a fine tune for the amount of backward scattering. Most of the atmospheric conditions can be probed by varying the value of the asymmetry parameter $g$: aerosols ($0.2 \leq g \leq 0.7$), haze ($0.7 \leq g \leq 0.8$), mist ($0.8 \leq g \leq 0.85$), fog ($0.85 \leq g \leq 0.9$) or rain ($0.9 \leq g \leq 1.0$)~\cite{Metari}. Changing $g$ from 0.2 to 1.0 increases greatly the probability of scattering in the very forward direction as it can be observed in Fig.~\ref{fig:HenyeyGreenstein}(left). Contrary to the Rayleigh scattering phase function, Henyey-Greenstein phase functions depict a strong forward peak directly linked to the asymmetry parameter $g$. This pronounced forward-scattering peak can easily be two orders of magnitude greater than at large scattering angles. Using the Ramsauer approach, the mean radius of the aerosol size distribution can be estimated from the width of the forward peak, for a fixed incident wavelength. Fig.~\ref{fig:HenyeyGreenstein} (right) shows the relation between $g$ values and mean particle radius for two different incoming wavelengths $\lambda$. Some of astroparticle projects measure this asymmetry parameter, giving a first estimation of the mean aerosol size present in the low part of the atmosphere (see Sect.~\ref{sec:exp_aerosol}).

\subsection{Multiple scattering by molecular and aerosol components}
\label{sec:mult_scattering}
Light coming from an isotropic source is scattered and / or absorbed by molecules and / or aerosols in the atmosphere. In the case of long distances or total optical depth values greater than about $0.5$, the single light scattering approximation -- when scattered light cannot be dispersed again to the detector and only direct light is recorded -- is not valid anymore. The multiple light scattering -- when photons are scattered several times before being detected -- has to be taken into account in the total signal recorded. Whereas the first phenomenon reduces the amount of light arriving at the detector, the latter increases the spatial blurring of the isotropic light source. Atmospheric blur is well-known for light propagation in the atmosphere and has been studied by many authors. A nice review of relevant findings in this research field can be read in~\cite{Kopeika_3}. Originally, these studies began with satellites imaging Earth where aerosol blur is considered as the main source of atmospheric blur~\cite{Dave,Pearce,Kopeika_1,Kopeika_2}. This effect is usually called {the adjacency effect}~\cite{Otterman,Tanre,Reinersman} since photons scattered by aerosols are recorded in pixels adjacent to where they should be. The problem of light scattering in the atmosphere has not analytical solutions. Even if analytical approximation solutions can be used in some cases~\cite{Kopeika_1,Ishimaru}, Monte Carlo simulations are usually used to study light propagation in the atmosphere. A multitude of Monte Carlo simulations have been developed in the past years, all yielding to similar conclusion: aerosol scattering is the main contribution to atmospheric blur, atmospheric turbulence being much less important. A significant source of atmospheric blur is especially aerosol scattering of light at near-forward angles~\cite{Kopeika_3,Reinersman}. The multiple scattering of light is affected by the optical thickness of the atmosphere, the aerosol size distribution and the height-dependent profiles of aerosol concentration.

The contribution of multiply scattered light to total light recorded depends not only on intrinsic properties of the atmosphere, but also on the source extent and on the integration time of the detector. Indeed, most scattered light arrives at the detector with a significant delay due to its detour. Thus, depending on the experiment and on the physics phenomenon probed, the multiply scattered light will not affect measurements with the same strength. In the case of very-high energy gamma rays and the detection of Cherenkov light beamed with respect to the air shower axis, scattered light contribution is insignificant since integration times are very short (typically much shorter than $100~$ns) and air showers are observed only at distances below one thousand meters due to the small field of view~\cite{bernholr}. Due to the short integration times, the near-forward scattering angles are responsible for most of the indirect light recorded by the camera, leading to a higher contribution of scattering on aerosols than scattering on molecular component. Contrary to Cherenkov radiation, fluorescence light is emitted isotropically and extensive air showers are mostly observed about ten kilometres from the telescope. Light emission is extended and integration times are typically few hundreds of nanoseconds. Therefore the reconstruction of the energy and the depth of maximum X$_{\rm max}$ of an extensive air shower is particularly affected by the multiply scattered light recorded by the fluorescence telescopes: it results in a systematic over-estimation of these two quantities if the multiple scattered fraction is not subtracted from the total signal. Four main studies about multiple scattering effect on air shower reconstruction have been done during the last ten years and are currently used in UHECR observatories: three of them based on Monte Carlo simulations~\cite{MS_roberts,MS_pekala,MS_louedec}, and the last one using only analytical calculations~\cite{MS_giller}. Contrary to analytical solutions, Monte Carlo simulations allow to follow each photon or photon packet emitted by an air shower and provide their number of scatterings during the propagation, and their arrival direction and time at the detector. All of these works predict the percentage of indirect light recorded at the detector within its time resolution (usually $100~$ns), for every shower geometry and aerosol conditions. However, only the parameterisation available in~\cite{MS_louedec} has explicitly a dependence on the asymmetry parameter $g$, a parameter directly linked to the forward scattering peak. This work was triggered by recent results obtained in the case of an isotropic point source and having demonstrated the importance of the asymmetry parameter on the point spread function of a ground-based detector~\cite{MS_louedec_PS,MS_pekala_PS}. An under-estimation of the aerosol size leads to a systematic over-estimation of the energy and the depth of maximum of the reconstructed air shower.

In the case of astronomical ground-based surveys, a particular attention is required to well estimate scattered light from extended sources such as starlight, zodiacal light or airglow in the Earth's atmosphere. Total zenithal optical depth values are usually assumed small, making possible analytical calculations in the single-scattering approximation~\cite{Wolstencroft,Staude,Bernstein,Noll}. However, this is not true anymore in the near-UV where zenithal optical depth values can be greater than $0.3$: a multiple scattering correction is usually applied by multiplying the single-scattered light by the factor $1 + 2.2\,\tau_{\rm m}(h,\theta=0^\circ,\lambda)$~\cite{Dave,Noll}. It takes into account only multiply scattered light by the molecular component, assuming a negligible effect from aerosols due to their low aerosol optical depth value. To our knowledge, this correction originally developed by J.V. Dave in 1964~\cite{Dave} has never been updated to consider aerosol contribution in cases where the aerosol component is not negligible. With the rise of astronomical all-sky surveys and the very low systematic uncertainties required, effect of these aerosols on multiply scattered light contribution should be investigated.

\subsection{Cloud extinction -- wavelength independent}
\label{sec:cloud_extinction}
In addition to atmospheric effects depending on the incident wavelength, data analysis in astroparticle physics experiments requires also recognition and correction for scattering and absorption of light by water droplets and ice crystals in clouds. Compared to aerosols found in the lowest part of the atmosphere, particles composing clouds are much larger in size, producing attenuation in the visible and NIR bands that is wavelength independent, the so-called {grey extinction}. Cloud cover is highly variable in both time and spatial direction. Clouds are located in the troposphere, the lowest part of the atmosphere extending up to about $20~$km in altitude. Clouds are usually categorised according to their base altitude range above Earth's surface: low (up to about $2~$km), middle (from $2~$km to $7~$km) and high (above about $7~$km). Low-cloud category includes mainly {cumulus} and {stratus}, mostly convective and non-convective, respectively. They can be also differentiated by the fact that cumulus are vertically extended and stratus are horizontally extended. Optically thick clouds (optical depth values greater than one) are mainly located in this part of the atmosphere. In contrast, {cirrus} which represent most of the highest clouds are generally non-convective and optically thin (optical depth values lower than $0.1$). Due to their corresponding altitude, they are made of ice crystals. Whereas thick cloud cover is usually easy to detect experimentally and to take into account their effect in data analysis, thin clouds as cirrus are much more difficult to be recognised. These clouds can affect recorded data, resulting in a systematic bias in analyses if their presence is not detected.

Depending on the astroparticle physics experiment, clouds have not the same effect on recorded data. In the case of ground-based astronomical surveys, it is crucial to estimate the amount and the structure of grey extinction on the recorded images. Indeed, cloud structure can be intricate with significant spatial variations across the field of view of the telescope, and may vary during the time interval of the exposure. Whereas optically thick clouds attenuate drastically the amount of light from astrophysical objects resulting in a useless survey, thin clouds cut just a part of the light. If this cloud attenuation is well evaluated, these corresponding surveys can be used for physics analyses, permitting to extend the telescope observing time. In such a correction, the spatial structure function of clouds has to be known since the light attenuation due to the cloud cover will not be similar over the whole field of view of the camera. This effect is especially important to correct for cirrus since their structure function is more complex. Such an effect, associated to the fact that clouds move in the field of view during the time exposure, is currently investigated in~\cite{GBlanc_private}. The problematic is not the same for the detection of extensive air showers. Clouds can either reduce the transmission of light from air showers or enhance the recorded light due to scattering in this over-density of matter. In these cases, clouds are easily detected and can be removed during the analysis. However, this is not anymore true for optically thin clouds which might be unnoticed in observations but could have an impact on the shape of the longitudinal profile and on the aperture of the detector, leading to wrong estimations for the energy spectrum of very-high energy gamma rays and ultra-high energy cosmic rays. To avoid as much as possible these errors and to optimise observing time, collaborations install \emph{in situ} auxiliary instruments to characterise the cloud cover above their observatories. The different techniques will be developed in Sect.~\ref{sec:exp_cloud}.

\section{The different instruments in atmospheric monitoring}
\label{sec:atmo_facilities}
Measurements of absorption and scattering properties of the atmosphere can exploit either natural sources to probe the atmosphere, or man-made illumination observing the atmosphere through backscattering. The atmospheric characteristics deduced are then used either in an event reconstruction software of the associated experiment or in a detailed atmospheric radiative transfer model as MODTRAN~\cite{MODTRAN_1,MODTRAN_2,MODTRAN_3}, a Monte Carlo simulation developed by the US Air Force Research Laboratory. This section enumerates the different facilities and indirect methods to probe molecular component, aerosol component and cloud cover above and in the surroundings of an astroparticle physics experiment. A main part of this section comes from a series of workshops discussing the atmospheric effects and how to estimate them in the case of astroparticle experiments~\cite{Workshop_1,Workshop_2,Workshop_3,Workshop_4,Workshop_5}, and spanning a decade of questionings and developments between the first workshop in 2003~\cite{Workshop_1} and the last one in 2014~\cite{Workshop_5}.

\subsection{Detection of the physics phenomenon itself used for atmospheric monitoring}
\label{sec:itself_monitoring}
Before describing the different facilities available to probe the atmosphere properties, collaborations in astroparticle physics have always developed some methods to deduce these properties directly from the measurement of the physics phenomenon studied, i.e.\ the air showers in the case of very-high gamma ray and ultra-high energy cosmic ray observatories, and celestial objects in the case of astronomical survey telescopes. However, they present systematic errors on atmospheric parameters greater than the ones obtained by a facility fully dedicated to atmospheric monitoring. If some of these methods are still employed to monitor the atmosphere, others became obsolete since their corresponding experiment requires nowadays a much better precision on their measurements.

This is exactly the case with this method that evaluates the aerosol optical depth using the fluorescence light emitted by extensive air showers themselves in the measurement of ultra-high energy cosmic rays. It has been developed by the HiRes collaboration and is based on the measurement of air showers in stereo, i.e.\ recorded by two fluorescence telescopes: the part of the shower viewed in common by two detectors at different distances permits to constrain the aerosol content in the atmosphere \cite{LRWiencke_stereo}. Assuming an atmosphere modelled only by molecular scattering and ozone absorption, the remaining attenuation attributed to aerosols can be determined. This technique requires no additional equipment and is insensitive to the absolute photometric calibration of the telescope camera. Its main limitation is statistical due to the low flux of cosmic rays in this energy range ($E \geqslant 10^{18}~$eV), making impossible an estimation of the aerosol attenuation on a hourly basis. The systematic uncertainty on the aerosol optical depth is about $0.014$, twice larger than the one obtained currently at the Pierre Auger Observatory with a technique using a dedicated laser (more details in Sect.~\ref{sec:exp_aerosol}).

Scientists using imaging air Cherenkov telescopes have also developed different techniques to calibrate their detector or estimate atmospheric conditions using only their instrument and air showers induced by very-high energy gamma rays. If they are in the surroundings of the telescope ($\sim 400~$m), high energy muons ($E \geqslant 8~$GeV) generated by air showers can be detected: when they pass throughout or close to a telescope, they emit a cone of Cherenkov light observed as a ring on the camera. The distribution of Cherenkov light in this ''muon ring'' is a function only of the muon distance from the telescope, atmospheric attenuation being assumed very small. Therefore this technique permits to obtain a conversion factor on the number of photoelectrons recorded by the detector for each Cherenkov photon hitting the telescope. This idea was firstly proposed by A.M. Hillas and J.R. Patterson~\cite{MuonRing_Hillas} in 1990, then applied to the Whipple telescope~\cite{MuonRing_1,MuonRing_2,MuonRing_3} and it is still the main method for absolute calibration of current imaging air Cherenkov telescopes~\cite{MuonRing_4,MuonRing_5,AtmoHEAD_Chalme}. A completely different method based on the trigger rate recorded by telescopes evaluates atmospheric effects from clouds or aerosols~\cite{HahnEtAl}. The maximum of the Cherenkov emission from air showers induced by very-high energy gamma rays is usually at an altitude from $5$ to $10~$km. Most of clouds and aerosol layers, located below this altitude range act as attenuators on Cherenkov light from the whole shower or part of it, resulting in a decrease of the trigger rate~\cite{FunkEtAl}. Whereas small clouds pass through the field of view of the telescope and reduce the trigger rate only on a short timescale (a few minutes), long-term atmospheric attenuators as aerosol layers or large cloud covers affect the trigger rate continuously. A new quantity based on this trigger rate, the so-called Cherenkov transparency coefficient, has been developed by the H.E.S.S. collaboration to characterise atmospheric attenuation during data acquisition. This coefficient is currently used only as a data quality parameter in the H.E.S.S. experiment, consisting in rejecting periods of data acquisition where the attenuation is too high. The next step would be to associate a correction factor to this coefficient in order to increase the duty cycle of the H.E.S.S. telescopes and to record events even in worse atmospheric conditions (as it is already the case in current ground-based cosmic ray observatories).

Concerning ground-based astronomical optical surveys, photometric data are usually calibrated using sets of standard stars whose brightness is known precisely from previous measurement campaigns. The first reference work is certainly the Landolt's catalogue providing magnitudes of several hundred stars near the celestial equator and measured with a photomultiplier tube on the Cerro Tololo 16 inch and the Cerro Tololo 1.5~m telescopes~\cite{Landolt_1,Landolt_2}. These relative photometric calibrations were achieved in five broad optical bandpasses, the so-called Johnson Kron Cousins photometric system \emph{UBVRI}, and reached an accuracy lower than $1\%$. Then, P.B. Stetson extended this catalogue to fainter magnitudes and released a catalogue of about $15\,000$ stars with $1-2\%$ accurate magnitudes~\cite{Stetson_1,Stetson_2}. Unfortunately, all these evaluations of brightness have been realised with a specific instrumental setup. Therefore a systematic uncertainty needs to be added when passing from the Landolt's system to the telescope considered in the analysis. To avoid this additional systematic uncertainty, the SDSS collaboration has designed its own photometric system based on the \emph{ugriz} bands~\cite{Fukugita1996}, extending the Landolt's and Stetson's works to fainter levels and increasing the number of studied stars to over one million~\cite{Ivezic_1}. This \emph{ugriz} photometric system is now widespread in the community and used now in photometric calibration efforts for other telescopes as SNLS~\cite{Regnault2009} or Pan-STARRS~\cite{Schlafly2012,Magnier2013}. All of them apply the same calibration algorithm, the so-called \"ubercalibration method, which simultaneously solves for the calibration parameters and relative stellar fluxes using overlapping observations~\cite{Padmanabhan_1}. This is a self-calibration method minimising the error dispersion in all observations and for all reference stars. It consists in separating the problem in a ''relative'' calibration -- establishing an internally consistent system -- and an ''absolute'' calibration -- providing the conversion factor between relative values recorded and physical fluxes. Finally, the ''absolute'' calibration is fully characterised with just a few parameters as the so-called zero point. It has been demonstrated that this method, combined with several observations of the same part of the sky, permits to obtain measurements of star brightness with an accuracy of about $1\%$. However, scientific programs of next imaging surveys -- composed of gigapixel CCD arrays forming a wide field of view -- will demand even more precise photometric calibrations to break through the $1\%$ barrier. Indeed, observing the sky at larger zenith angles will change drastically the depth of atmosphere between the telescope and the source. This basic fact indicts that variations of atmospheric attenuation should be one of the main limitations to the precision of the next ground-based all-sky surveys. The main idea to improve still the accuracy of next ground-based photometric measurements consists in separating the instrument and the atmosphere explicitly in calibrations~\cite{Stubbs_2}. Indeed, for now, unmodelled variations of the atmosphere are responsible for almost all the calibration error budget. The concept would be to directly measure the atmospheric transmission using an instrumentation dedicated to this task as an auxiliary ground-based telescope or specific instruments commonly used in atmospheric sciences.

Through these three examples, it is interesting to observe how instruments fully dedicated to the atmosphere are more and more included in astroparticle experiments. With the increasingly demand on precision, variations of the atmosphere become the limiting factor. Whereas experiments in ultra-high energy cosmic rays have already developed all this specific atmospheric monitoring by installing weather stations, IR cameras for cloud detection, LiDARs for aerosol detection or using global atmospheric models, collaborations in imaging air Cherenkov telescopes or ground-based astronomical surveys are just designing the optimised procedures to well characterise the atmosphere. These works would certainly lead to install \emph{in situ} specific instruments developed and used in atmospheric sciences since their current telescopes provide only a global estimation of atmospheric effects with no information about vertical structure of the atmosphere. The two main goals will be to improve the precision on measurements -- smaller systematic uncertainties -- and to relax the quality cuts on data -- a higher duty cycle of the telescope data taking. The purpose of the next sub-sections is to present these different instruments.

\subsection{Molecular component}
\label{sec:exp_molecular}
As explained in Sect.~\ref{sec:molecular_absorption} and Sect.~\ref{sec:molecular_scattering}, the molecular component is characterised by its well-mixed gases (dinitrogren, dioxygen, etc) and variable gases as water vapour or ozone present in much lower quantity. Whereas permanent gases fix the general properties of the atmosphere as its height-dependent profiles of state variables, variable gases follow different horizontal and vertical spatial distributions in the atmosphere, requiring a dedicated monitoring~\cite{EJP_MWill}. Weather stations are used to follow the evolution of atmospheric state variables at ground level. They are usually powered by solar panels and composed of temperature, pressure, humidity and wind (speed and direction) sensors, transmitting measured values every a few seconds. Precisions on measurements are usually about $0.5^\circ$C for temperature, $0.5~$hPa for pressure and about $2\%$ for relative humidity. To reduce as much as possible systematic uncertainties in measurements, ground-based weather stations have to be placed correctly. For instance, concerning wind measurements, the best way is to install an anemometer in the top of a $10~$m mast in order to respect international standards. In addition to ground-based weather stations, measurements of the height-dependent profiles of state variables are needed to well estimate atmospheric effects up to the top of the atmosphere. The most widespread technique is to use helium-filled weather balloons to launch meteorological radiosondes, providing values of temperature, pressure, relative humidity and wind speed every about $20~$m from ground level to about $30~$km. Measurement errors are similar to what we have with weather stations, except for the relative humidity where they are slightly larger. Meteorological data and GPS position are sent continuously during the flight to a station located at ground. Whereas horizontal directions of the balloon are mainly governed by the wind, vertical movement follow the balloon buoyancy.

\begin{figure*}[!t]
\centering
\resizebox{0.55\textwidth}{!}{%
\includegraphics{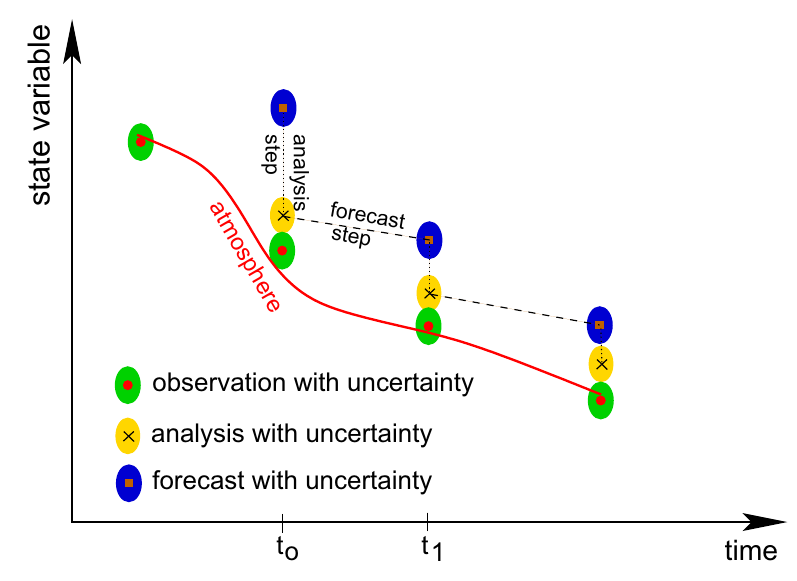}
}
\hfill{}
\resizebox{0.42\textwidth}{!}{%
\includegraphics{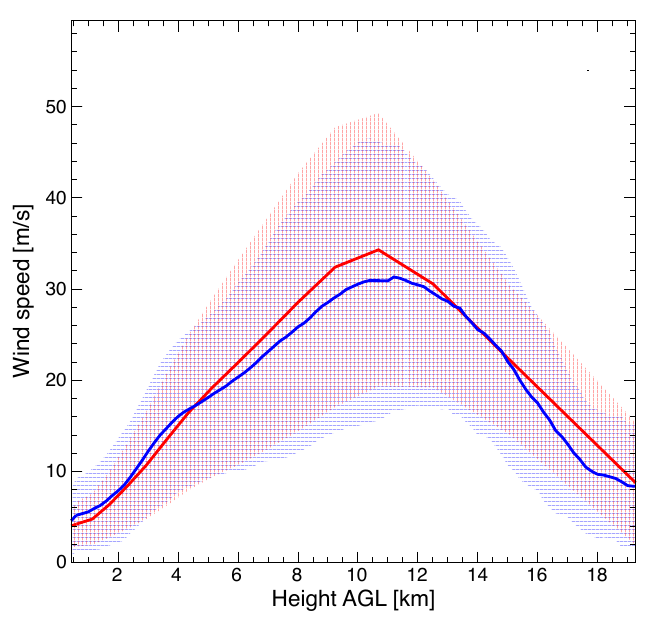}
}
\caption{(left) Sketch of the data assimilation process with observation, analysis and forecast values (from~\cite{Auger_GDAS}). At a given time $t_{\rm 0}$, observations provide the value of a state variable. The analysis step combines observation and forecast to describe the current state in a better way than the forecast. This obtained value is the initial point for an expected value for a later time $t_{\rm 1}$. (right) Comparison between radiosonde data and GDAS data at the Pierre Auger Observatory. Continuous lines represent the averaged wind profile for balloon data ({in blue}) and GDAS data ({in red}) (from~\cite{HYSPLIT_ECRS}).}
\label{fig:instru_molecular}
\end{figure*}

All these local measurements show that it can exist large daily fluctuations in temperature, pressure or humidity (and wind speed). But operations associated to radio soundings represent a large burden, both in terms of funds and manpower. A possibility to avoid this charge is to use data from global atmospheric models. The latter are based on data assimilation, i.e.\ a technique used in numerical weather prediction where calculations take into account the real-time conditions of the atmosphere as boundary condition. Atmospheric models provide the atmospheric state at a given time and at a given position on a latitude/longitude grid (as a good approximation, horizontal uniformity of state variables can be assumed if the Earth's surface is more or less plane). For a given position on the grid, values of a state variable are given at different constant pressure levels. It consists in collecting all the available meteorological data from weather stations, meteorological balloons, satellites, aircrafts, etc. Then, for a given time, the value of a state variable is know from observations, but also predicted by the atmospheric model: the data assimilation consists in combining observations and forecasts to estimate a 3-dimensional image of the atmosphere~\cite{MWill_thesis}. This algorithm is then repeated for a later time. A sketch illustrating the concept of data assimilation is given in Fig.~\ref{fig:instru_molecular}(left). The main meteorological data assimilation projects around the world are ERA-Interim developed by ECMWF (European Centre for Medium-range Weather Forecasts) \cite{ECMWF_website}, GDAS by NCEP (National Centres for Environmental Prediction) \cite{NCEP_website} and GOES-5 by GMAO (Global Modelling and Assimilation Office) \cite{GMAO_website}. An analysis conducted by the Pierre Auger collaboration has validated the GDAS data when compared to local measurements made at the observatory, and has demonstrated that air shower reconstruction was improved by incorporating GDAS data in the process instead of using spare measurements operated at the observatory~\cite{Auger_GDAS,HYSPLIT_ECRS} (see Fig.~\ref{fig:instru_molecular}(right)). Inspired by these results, the JEM-EUSO collaboration is investigating the possibility to apply a similar way to get atmospheric state variables~\cite{ICRC13_JEMEUSOatmo}. Whereas experiments in ultra-high energy cosmic rays are already familiar with these global atmospheric models, systematic uncertainties required up to now in ground-based gamma astronomy or ground-based astronomical surveys have not needed such models. However, with the new goals in precision currently discussed \cite{CTA,Ivezic_1}, this practice would certainly change in a near future and applying these models seems to be a natural solution.

Even if precipitable water vapour can be extracted from global atmospheric models or meteorological balloon radio soundings, it would lead to large systematic errors since it is highly variable in space and time with fluctuations of $5$ to $10\%$ per hour possible. Yet water vapour remains one of the most poorly characterised meteorological parameters. As already said in Sect.~\ref{sec:molecular_absorption}, water vapour is not a well-mixed atmospheric constituent and its value measured at ground level does not reproduce the behaviour of its total column height. In the last thirty years, two additional remote sensing methods to retrieve water vapour continuously and automatically have become available. The first method is based on the radiometry of air molecules in the $10-200~$GHz spectral range and especially the thermal emission of water vapour near the $22.235~$GHz ($1.35~$cm) spectral line~\cite{Dicke1946,England1992}. Indeed, in addition to absorption lines in optical wavelength range, water molecules absorb also electromagnetic waves in the microwave and radio domains. It consists in measuring electromagnetic emission from transitions between different states of rotational energy. The line width of these observed spectral emission lines is affected by different broadening processes, the dominating one in the low part of the atmosphere being the pressure broadening produced by collisions between the target molecules (in this case, H$_{\rm 2}$O) and other air molecules. Since it exists a relation between pressure and altitude, pressure broadening of emission lines is used to estimate the altitude of probed molecules through inversion methods, with a resolution of $5-10~$km~\cite{GerberEtAl}. Usually, water vapour radiometers are fully steerable in both azimuth and elevation, providing a full sky coverage. Ground-based microwave radiometers are able to operate continuously for the retrieval precipitable water vapour with a high temporal resolution, but providing measurements not reliable during rainfall. The second method is based on the Global Positioning System (GPS). It allows derivation of the water vapour path as well as its vertical distribution from the path delay of the GPS signals~\cite{Businger,Ware1996}. Microwave radio signals transmitted by GPS satellites to Earth-based receivers are delayed (refracted) by the atmosphere. Part of this delay is due to the presence of water vapour, its corresponding value being nearly proportional to the quantity of water vapour integrated along the signal path~\cite{Rocken,Borbas}. Given the development of GPS satellites, a basic Earth-based GPS receiver permits to monitor the distribution of water vapour with a large time coverage and makes it a solution that can be considered.

Among variable gases present in the atmosphere, ozone is probably the most fluctuating in concentration and spatial distribution after water vapour. Since ozone is a key variable needed for understanding climate processes and change, the number of instruments to measure it was intensively increased during the last decades. Total ozone column is measured from ground using Dobson or Brewer spectrophotometers. They record ultraviolet light from the sun at two to six different wavelengths from $305$ to $345~$nm. The measuring principle uses the fact that ozone absorption depends on the wavelength: whereas light is strongly absorbed at $305~$nm, this is not anymore the case at $325~$nm. The ratio between the two recorded light signals gives a direct estimation of column ozone in the light path from the sun to the spectrophotometer. The Brewer instrument is based on the same measuring principle but using five different wavelengths between $306$ and $320~$nm. Regarding the ozone vertical profile, it is measured using ozone sondes embedded in weather balloons and probing the vertical structure of the atmosphere from ground to about $30~$km. Ozone sondes are composed of a pump that pulls air into a chamber with potassium iodine, producing a chemical reaction converting the potassium iodine into iodine. Information about potassium iodine reaction with ozone is transmitted via radio waves to the ground. The corresponding vertical resolution is about $200-300~$m due to the response time of the electrochemical sensor. Even if it is less widespread, ozone profile can also be measured using LiDAR and microwave instruments, with a particular interest in the stratosphere and mesosphere. Indeed, they typically cover the altitude ranges $10-50~$km and $20-70~$km, respectively, with associated vertical resolutions of about $100-200~$m and $5-10~$km.

Finally, reactive gases as carbon monoxide CO, volatile organic compounds VOCs, oxidised nitrogen compounds NOx or sulphur dioxide SO$_{\rm 2}$ have to be monitored. All of them play an important role in the chemistry of the atmosphere concerning climate or the formation of aerosols. Each component requires a specific procedure and setup to measure it. Since their concentration is usually very low and depends strongly on the location, explaining in detail the procedure for each chemical component is beyond the scope of this review. We refer the reader to different reports to get further information on experimental procedures concerning each component: carbon monoxide~\cite{CO}, volatile organic compounds~\cite{VOCs} or oxidised nitrogen compounds~\cite{NOx}. These procedures are usually practiced on two types of monitoring platforms: \emph{in situ} monitoring at atmospheric observatories allowing for long-term and frequent sampling, or mobile platforms as aircrafts, ships or trains providing unique opportunities to probe the horizontal and vertical distributions of chemical components.

\subsection{Aerosol component}
\label{sec:exp_aerosol}
Atmospheric aerosols play an important role in climate change or air quality. Aerosols have many possible sources as sea spray, mineral dust, or chemical reactions of gases in the atmosphere. Complexity of these mechanisms is so great that it leads to large uncertainties in our quantitative understanding of the aerosol role in climate change or air quality. Therefore it is not surprising to see that a large range of detectors has been developed during the past decades to measure atmospheric aerosols, i.e.\ their concentration, their shape, their size, their chemical composition, etc. Whereas sampling techniques are the best way to characterise aerosols, they probe only a small volume of atmosphere and are often limited in statistics -- complete reviews of these methods have been written by J.C. Chow~\cite{Chow1995} and P.H. McMurry~\cite{McMurry_review}. Thus, \emph{in situ} measurements of radiative properties are usually a good opportunity to estimate indirectly and continuously aerosol properties. The rest of this subsection presents briefly most of these techniques.

\subsubsection{Aerosol sampling techniques: mass concentration, size distribution and chemical composition}
\label{sec:aerosol_sampling}
Aerosol measurements in sampling techniques can be categorised following the S.K. Friedlander's suggestion \cite{Friedlander_1,Friedlander_2}: measurements providing a single piece of information integrated over size and composition, and those giving more detailed resolution with respect to size and time. In both categories, the design of aerosol sampling inlet requires a careful consideration. The purpose of the inlet is to provide an aerosol sample representative of ambient air, i.e.\ a system minimising local influences, or having an aerosol transmission efficiency that does not vary with wind direction or wind speed. In other words, the ideal inlet would collect 100\% of aerosols in a specified size range. As already said in Sect.~\ref{sec:aerosol_scattering}, aerosols are hygroscopic, especially in nucleation and accumulation modes: water typically constitutes more than half of these aerosol modes at relative humidity greater than roughly $80\%$. Humidity control and size cuts are the best ways not to get aerosol data biased by water. Filter samplers are often used to store aerosols in the aim to analyse them later in laboratory, remaining the most robust method up to now. Since final results are expressed in terms of air concentration, air volume for each aerosol sample is also determined by integrating airflow rate over the sampling duration. This duration varies with locations, sampling rates or analytical sensitivities but typically ranges from several hours to a day or more under clean atmospheric conditions.

Instruments integrating aerosols over a given size range are often used for their simplicity. Mass concentration of aerosols is a fundamental parameter: the international air quality standards require measurement of mass concentration of particles smaller than $2.5~\upmu$m (PM$_{\rm 2.5}$) or $10~\upmu$m (PM$_{\rm 10}$) aerodynamic diameter\footnote{In general, particles have irregular shapes with actual geometric diameters that are difficult to estimate. \emph{Aerodynamic diameter} is an expression of a particle aerodynamic behaviour as if it were a perfect sphere with unit-density and diameter equal to the aerodynamic diameter.}. Its measurement is often done gravimetrically, where it is determined from the net aerosol mass on a filter, divided by the volume of air sampled. During this estimation, relative humidity and temperature need to be fixed at reference values not to bias comparison with other aerosol samples. Analytical precisions for gravimetric analyses are currently about $\pm 1~\upmu$g. Limitations in filter measurements are gas adsorption on substrates (typically organic gases on quartz filters~\cite{Cadle1983}), evaporation of semi-volatile components~\cite{Turpin1993}, and chemical reactions between collected particles and substrates \cite{McMurry1989,Smith1978}. To avoid these limitations and to reduce the manpower charge, different automated techniques for continuous or semi-continuous aerosol mass concentration measurements have been developed, as beta-meters, piezoelectric crystals or harmonic oscillating elements~\cite{Baltensperger2001}. While there are obvious advantages of employing these automated instruments, there are still some issues with using these instruments for long-term measurements.

Aerosol size distribution is made of several modes, ranging from a few nanometres to a few tens of microns. These different aerosol modes have not the same origin or the same chemical composition. Therefore size-resolved measurements are useful to understand the behaviour of specific aerosol size ranges. The most widespread type of instruments is the single-particle optical counters (also called aerosol spectrometers) measuring in real time the amount of light scattered by an individual particle when it traverses a tight focused beam of light. A fraction of the scattered light is recorded by a photo-detector and converted into a voltage pulse. This is the amplitude of this pulse that estimates the particle size, using a calibration curve obtained from measurements of spherical particles of known size and composition. Even if they are commonly used nowadays in aerosol studies since they are cheap and easy to use, some limitations remain: they tend to heat aerosols leading to a systematic smaller size for hygroscopic aerosols, their calibration curve is obtained for a specific chemical composition which is not always representative of aerosols probed, and aerosols with irregular shapes will false their size estimation. Another kind of instruments more sophisticated permits to solve these issues, especially giving valuable information about the shape and/or refractive index of atmospheric particles: the multi-angle aerosol spectrometer probe (MASP) \cite{Baumgardner1993}, measuring light scattered by individual particles for polar angles of $30^\circ-60^\circ$ and $120^\circ-150^\circ$. Other techniques based on the aerodynamic particle size~\cite{WilsonLiu1980} (sizes greater than $0.2-0.5~\upmu$m), particle electrical mobility~\cite{Whitby1966,Liu1974,Keady1983} ($3-900~$nm), particle diffusivity~\cite{Knutson1988,Knutson1995} (particles smaller than $0.1~\upmu$m) or particle growth by condensation~\cite{Stolzenburg1991} ($3-10~$nm) are also available to estimate aerosol size distribution. 

Concerning chemical composition of aerosols, this analysis is most of the time done in laboratory. Two aerosol categories are mainly analysed wit this procedure: ionic species and mineral dust. Ionic species including sulphate, nitrate, chloride, sodium, ammonium, potassium, magnesium or calcium represent a major part of aerosol mass. Their presence is usually evaluated applying ion chromatography to aerosol filter samples: sulfate is the most studied chemical element and ubiquitous in aerosols, nitrate is mainly produced by reaction of nitric acid vapour with alkaline components in aerosols, and sea salt ionic components dominate the mass of the coarse mode over oceans and coastal areas. Sea salt components, as mineral dust components (aluminium, silicium, iron, titanium, scandium) and trace components (nickel, copper, zinc, lead), can be also analysed by destructive methods (atomic absorption spectroscopy (AAS) or inductively coupled plasma mass spectroscopy (ICPMS)) or non-destructive methods such as instrumental neutron activation analysis (INAA), proton induced X-ray emission (PIXE), X-ray fluorescence analysis (XRF) or scanning electron microscope (SEM) equipped with an energy dispersive X-ray system (EDX). The latter is the most widespread technique for individual particle analysis, providing particle morphology and elemental composition for atomic numbers greater than 11 (sodium Na). The main limitation for this technique is that obtaining data with an enough statistical significance becomes considerably time consuming. The EDX system is used to avoid volatilisation from aerosol samples when they are exposed to vacuum conditions and are heated by the electron beam. Even if real-time measurements of chemical composition are available, they are still in development. The most advanced techniques concern particulate carbon analysers, and particulate sulfur and nitrogen species analysers (see~\cite{McMurry_review} for more details).

Even if all these aerosol sampling techniques are very well-known in atmospheric sciences, this is not the truth in astroparticle physics yet. To our knowledge, aerosol sampling has been operated only at the Pierre Auger Observatory during one half a year where aerosols have been collected using filters. Then PIXE and SEM/EDX techniques have been applied in laboratory to get the chemical composition of aerosols~\cite{MIM_NIM}. Aerosol sampling have continued one year later using this time an aerosol spectrometer to estimate precisely the aerosol size distribution~\cite{MIM_ICRC13}. All these measurements are an opportunity to understand better the origin of aerosols present at the observatory.

\subsubsection{Measurement of aerosol radiative properties}
\label{sec:aerosol_radiative}
Sampling techniques to measure aerosols are numerous. Even if they provide a precise characterisation of atmospheric particles, they probe only the air volume just around the detector and do not inform us about height-dependent aerosol properties. Last but least, some of them as filter samplings require manpower. An alternative to these sampling techniques is to estimate directly aerosol radiative properties. Indeed, in the case of spherical particles at least, aerosol radiative properties are linked to aerosol properties via the Mie scattering theory. Moreover, this method does not seem incongruous when we know that astroparticle physics experiments discussed in this review record light produced by extensive air showers or coming from celestial objects. Instruments measuring aerosol radiative properties can be divided into two categories: ''passive'' techniques exploiting natural sources to probe the atmosphere, and ''active'' techniques recording light produced by an associated laser. Whereas the former are cheapest, only the latter provide a full description of the height-dependent aerosol properties. The purpose of this subsection is to list briefly the different instruments with their associated deliverables and limitations.

Sun-photometers are probably the most common instrument in atmospheric sciences to monitor \emph{in situ} column integrated aerosol optical properties in the category of ''passive'' techniques. It consists in pointing the sun through the day thanks to a tracking system, and measuring solar radiation at different spectral bands within the visible and near-infrared spectrum. Since original solar radiance is well-known, differences observed in sun-photometer measurements are due to the atmosphere. For each spectral band studied, it is possible to estimate the total aerosol optical depth (AOD) using the Beer-Lambert law (see Eq.~\eqref{eq:beer}), i.e.\ the total extinction of solar radiation by aerosol scattering and absorption between the top of the atmosphere and the ground-based detector. Also, simultaneous AOD measurements at several wavelengths permit to estimate the Angstr\"om coefficient $\gamma$ which gives an indication of the aerosol size distribution. Since ozone or water vapour have specific absorption bands in the atmospheric transmission spectrum, measurement at one of these specific wavelengths permits to constrain the total atmospheric column of these constituents. In a less direct way, it is possible to retrieve inversion aerosol products as the single scattering albedo or the aerosol phase function from almucantar scans of radiance combined to inversion algorithms~\cite{Dubovik}. However, sun-photometers cannot be operated during nights, i.e.\ exactly the periods where astroparticle physics experiments record data. A work is currently in progress to replace the sun by the moon, the difficulty being that the variation of the moon illumination is inherent to the lunar cycle~\cite{Lunarphotometer}. Once this prototype would be validated, it should increase drastically our knowledge on aerosols during nighttime. Another solution consisting in observing stars during nighttime is also investigated by several groups: following standard stars during their path in the sky via a tracking mode, it is possible to measure their luminosity and to estimate the atmosphere transparency by inversion algorithms. This method is similar in a simplified approach to techniques applied in ground-based astronomical survey telescopes to monitor the atmosphere using a star catalogue. As in the case of sun-photometers, they record signals from $300$ to about $800~$nm with several filters centred at different wavelengths and estimate the AOD value and the Angstr\"om coefficient through the night. The two main instruments exploiting this idea are the UVscope instrument based on a multi anode photomultiplier tube~\cite{UVscope,UVSiPM} and the F/(Ph)otometric Robotic Atmospheric Monitor (FRAM) based on a Cassegrain-type telescope coupled to a CCD camera~\cite{FRAM}. Preliminary and promising results are already available for these two facilities. We mention here also a work in progress to estimate aerosols using an all-sky scanning infrared radiometer: even if this type of instruments is first dedicated to detect clouds, monitoring of the aerosol component seems to be possible~\cite{AtmoHEAD_Daniel}. Indeed, large aerosols as pollen or sand grains are expected to contribute up to $30~$W/m$^2$ to the sky radiance in the $8-13~\upmu$m atmospheric window~\cite{Dalrymple}. Obviously, sun / star / lunar-photometry is better suited for the study of widespread hazy conditions than for the study of smoke plumes. A smoke plume tends to be very dense and is very localised. In the case of large aerosol size, typically greater than roughly $500~\upmu$m, an X-band Doppler radar system can be used to measure the terminal settling velocities of aerosol particles as volcanic ashes or water droplets. The famous model named Pludix is first dedicated to the characterisation of rainfalls within a sampling volume surrounding it~\cite{Scollo,Bonadonna,AtmoHEAD_Leto}. Falling objects crossing the antenna beam ($\lambda = 9.5~$GHz) generate power echoes backscattered to the radar with a frequency shift related to the object velocity. The received signal is then analysed with different algorithms to obtain an aerosol size distribution into 21 bands of mean diameter between $0.8$ and $7.0~$mm~\cite{Prodi2000}. The main limitation for this technique is that chemical composition of particles studied needs to be known in the analysis algorithms.

Other techniques based on the measurement of the aerosol phase function exist to estimate the aerosol size at ground. The most famous is the integrating nephelometre available commercially~\cite{BeutellBrewer1949,HeintzenbergCharlson1996}. It consists in illuminating a volume of air with a diffuse light source, at one or several wavelengths depending on the instrument model. A photon-counting detector with its axis perpendicular to the light source records a part of the scattered light from this illuminated air volume. Ranges of angular integration are typically $7^\circ-170^\circ$ for the ''total'' scatter coefficient and $90^\circ-170^\circ$ for the ''back'' scatter coefficient. Even if instruments estimating directly the asymmetry parameter $g$ do not exist, the ratio of these two scatter coefficients can be linked to the $g$ parameter~\cite{Marshall1995,WiscombeGrams1976}, giving an easy and cheap opportunity to estimate roughly the aerosol phase function. Collaborations in ultra-high energy cosmic rays have developed also a technique based on the measurement of the aerosol phase function to estimate the aerosol size~\cite{HiRes_CLF,BenZvi_APF}.  Aerosol phase function monitors, in conjunction with UV telescopes, are used to measure the asymmetry parameter $g$ on an hourly basis during data acquisition. The light sources emit a near-horizontal pulsed light beam in the field of view of their nearby UV telescope. Each monitor contains a collimated Xenon flash lamp source, firing an hourly sequence of $350~$nm and $390~$nm shots. The aerosol phase function is then reconstructed from the intensity of the light observed by the UV telescope as a function of the scattering angle $\zeta$, for angles between $30^{\circ}$ and $150^{\circ}$. After corrections for geometry, attenuation and collection efficiency of each pixel, the binned signal $S(\zeta)$ observed is subjected to a 4-parameter fit
\begin{equation}
S(\zeta) = C\, \left[\alpha_{\rm m}(h_{\rm gl})\, P_{\rm m}(\zeta) + \alpha_{\rm a}(h_{\rm gl})\, P_{\rm a}(\zeta |g,f)   \right],
\label{eq:apf_monitor}
\end{equation}
where $\lbrace C\,\alpha_{\rm m}(h_{\rm gl}),C\,\alpha_{\rm a}(h_{\rm gl}),g,f\rbrace$ are the fit parameters \cite{BenZvi_APF}. The first two fit parameters can be used to estimate the molecular extinction and the aerosol extinction, respectively; while $g$ and $f$ are used to estimate the aerosol size distribution. Recently, a new method based on very inclined laser shots fired by a steerable system and recorded by a UV telescope was also developed and is still in progress~\cite{KLouedec_thesis,MyICRC}.

All the instruments presented up to now provide no information about the height-dependence of aerosol properties. But it is recognised that measuring vertical profile of aerosols is a natural complement to total column aerosol observations made by ground-based sun photometers or satellites (see Sect.~\ref{sec:exp_satellite_networks}). The aim is to identify aerosol layers, aerosol optical properties -- backscatter and extinction coefficients at given wavelengths, Angstr\"om coefficient --, and aerosol microphysical properties -- concentration, size distribution, refractive index. Ground-based laser facilities can monitor continuously the structure of the planetary boundary layer (the lowest part of the atmosphere), its height and its variability with time (e.g.\ diurnal mixing). Also, retrieval of microphysical properties for elevated aerosol layers is an important point regarding the development of extensive showers in the atmosphere and is feasible only for advanced ground-based laser facilities. The LiDAR (Light Detection And Ranging) technique consists in emitting pulses of light up through the atmosphere and in recording light scattered back by an optical receiver on the ground as a function of time. The difference in light-travel time to different altitudes provides a method for probing the vertical structure of the atmosphere~\cite{LIDAR}. Aerosol optical properties can also be obtained using multi-elevation-angle measurements by a scanning LiDAR, allowing more accurate estimation of vertical and horizontal spatial extensions~\cite{BenZvi_LIDAR,EJP_VRizi}. Light pulses emitted into the atmosphere can be scattered elastically -- light re-emitted at the same wavelength $\lambda_{\rm em}$ -- or inelastically -- light re-emitted with a wavelength shift $|\lambda_{\rm em} - \lambda_{\rm rec}|$ due to excitement of internal degrees of freedom in the scattering particle. Since inelastic scattering cross section is much smaller than elastic cross section, more intense light source, longer detection time and larger detector aperture are necessary. The LiDAR return signal $S(r,\lambda_{\rm em})$ is given by the so-called LiDAR equation
\begin{equation}
\begin{split}
&S(r,\lambda_{\rm em}) = P_{\rm 0}\,\frac{c\,\Delta t}{2}\,A\,\frac{G(r)}{r^2}\,\beta(r,\lambda_{\rm em})\,T^\uparrow(r,\lambda_{\rm em})\,T^\downarrow(r,\lambda_{\rm rec}) \\
&\underset{\lambda_{\rm em} = \lambda_{\rm rec} = \lambda}{\overset{\rm elastic}{\longrightarrow}} P_{\rm 0}\,\frac{c\,\Delta t}{2}\,A\,\frac{G(r)}{r^2}\,\beta(r,\lambda_{\rm em})\,{\rm e}^{-2\tau(r,\lambda_{\rm em})},
\end{split}
\label{eq:LIDAR}
\end{equation}
where $r$ is the scattered photon distance from the optical receiver, $\Delta t$ the laser pulse duration, $A$ the collection area of the telescope, $G(r)$ the overlap factor between the telescope and the laser cone (equal to one in the ideal case), $T^\uparrow$ and $T^\downarrow$ the total atmospheric transmission factors, and $\beta(r,\lambda_{\rm em})$ the total backscatter coefficient with $\beta = \beta_{\rm a} + \beta_{\rm m}$ (a: aerosol / m: molecular). In the case of elastic scattering only, $\lambda_{\rm em} = \lambda_{\rm rec}$, and the LiDAR equation is reduced to the second part of Eq.~\eqref{eq:LIDAR}, where $\tau(r,\lambda_{\rm em})$ is the total optical depth equal to $\int_0^r \alpha(\lambda_{\rm em},r') {\rm d}r'$ with $\alpha(\lambda_{\rm em},r)$ the total extinction coefficient ($\alpha = \alpha_{\rm a} + \alpha_{\rm m}$). We distinguish between LiDAR systems detecting only elastically scattered light from both aerosols and molecules, called elastic-backscatter LiDARs~\cite{LIDAR,AtmoHEAD_Bourgeat}, and those detecting the molecular scattering separately from the aerosol scattering thanks to backscattered light from roto-vibrational excitation of atmospheric molecules (N$_{\rm 2}$ / O$_{\rm 2}$), called Raman LiDARs~\cite{LIDAR_Raman}, or via Rayleigh scattering with a high spectral resolution LiDAR (HSRL)~\cite{LIDAR_HSRL}. In every technique, instruments can be operated at multiple wavelengths simultaneously. Accurate retrieval of extinction $\alpha(\lambda,r)$ and backscatter $\beta(\lambda,r)$ profiles without making assumptions on the aerosol properties is only possible with the measurement of two independent signals.

The LiDAR system with the lowest complexity and the most widespread is the elastic-backscatter one measuring the backscatter signal $S(r,\lambda)$ at one wavelength. In its basic form, an elastic-backscatter LiDAR is called a ceilometer, an optical facility that can be purchased. Elastic-backscatter LiDARs are useful to probe the vertical structure of the atmosphere. A great interest concerns the height of the planetary boundary layer which can be estimated if the overlap factor of the system $G(r)$ is known. After having estimated the molecular parameters $\{\beta_{\rm m}(\lambda,r)$, $\alpha_{\rm m}(\lambda,r)\}$ using weather radio soundings or global atmospheric models, it remains two unknowns to be determined in Eq.~\eqref{eq:LIDAR} leading to an undetermined system: the aerosol backscatter coefficient $\beta_{\rm a}(\lambda,r)$ and the aerosol extinction coefficient $\alpha_{\rm a}(\lambda,r)$. Inversion algorithms assuming a typical LiDAR-ratio profile ($\alpha_{\rm a}(r)/\beta_{\rm a}(r)$) are then applied to estimate the aerosol backscatter coefficient~\cite{Klett,Fernald}. In most cases, they provide an aerosol backscatter coefficient with an associated error of 10\% and a quite uncertain aerosol extinction coefficient with a typical error of 50\%. These errors reach their highest values at short wavelengths. To reduce the uncertainty on the aerosol extinction coefficient to $10\%$, more complex LiDAR facilities consisting in measuring two signal profiles are required: one channel records the total backscattered signal and the second channel a pure molecular backscattered signal, i.e.\ without any need of information about the molecular density profile from weather radio soundings or global atmospheric models. Thus, the profiles of the aerosol backscatter coefficient $\beta_{\rm a}(\lambda,r)$ and of the aerosol extinction coefficient $\alpha_{\rm a}(\lambda,r)$ can be determined independently from each other, and the LiDAR-ratio profile is directly deduced. Raman LiDARs are based on the vibrational or rotational Raman scattering from nitrogen or oxygen. Even if rotational Raman scattering has a cross section about thirty higher than vibrational Raman scattering, the latter is usually employed since the wavelength shift is larger, i.e.\ easier to detect. If the Raman LiDAR is operated during daytime, a filter with a width of a few tenths of nanometre has to be added, reducing considerably the light collection efficiency. Regarding the HSR LiDARs, they can be operated equivalently at day and night. They are based on the separation of the molecular scattering from the aerosol scattering using the Doppler frequency shift produced when photons are scattered in random thermal motion. Whereas molecular velocities are described by a Maxwellian distribution with an associated width of about $300~$m/s, aerosols move with velocities fixed by the wind ($\sim 10~$m/s) and turbulence ($\sim 1~$m/s). The resulting frequency distribution of light backscattered from the atmosphere is a narrow spike near the frequency of the laser caused by aerosols riding on a much broader distribution produced by the molecular components. The use of an ultra-narrowband filter permits to isolate the two scattering origins~\cite{Eloranta_livre}. Even if this technique gives better results than Raman LiDARs in theory, it involves a much more complex system to develop and to maintain. Therefore a Raman LiDAR is preferred over an HSR LiDAR. For instance, several Raman LiDARs with different designs are now in construction to fulfil the requirements of measurement precision for the next ground-based Cherenkov telescope array~\cite{AtmoHEAD_Doro,ICRC_Doro,AtmoHEAD_Pallotta,ICRC_Arg,ICRC_Raman_IFAE_LUPM}. In all LiDAR techniques listed above, they can be used at multiple wavelengths, supplemented by polarisation channels or coupled to a sun-photometer to obtain a better estimation of particle microphysical properties: the distinction of clouds from aerosol layers is a possible application (see Sect.~\ref{sec:cloud_lidar} for further details).

\begin{figure}[!t]
\centering
\resizebox{.5\textwidth}{!}{%
\includegraphics{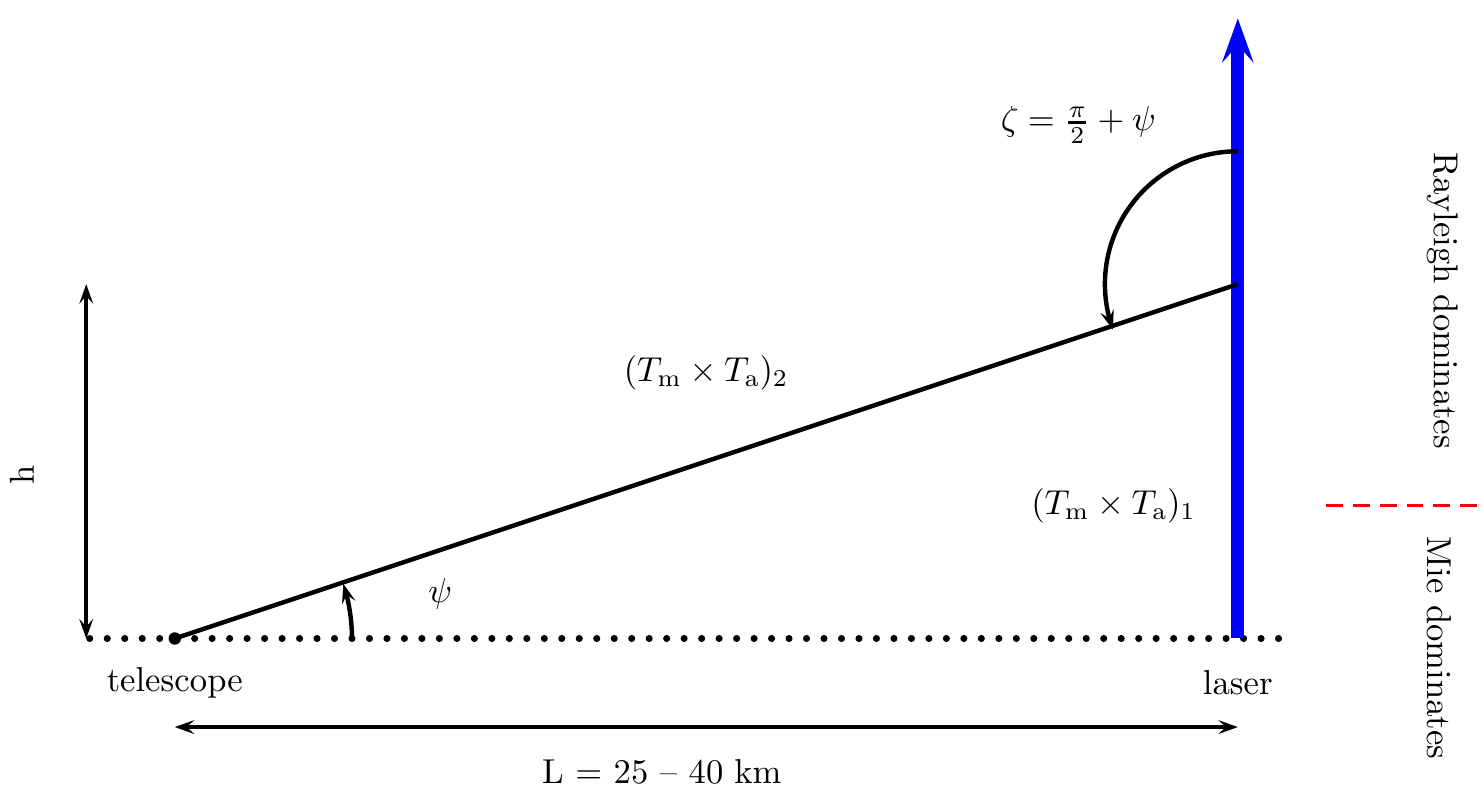}
}
\caption{{Geometrical arrangement, viewed from the side, of the laser facility and the telescope.} The light is scattered out of the laser beam at a height $h$ corresponding to an elevation angle $\psi$ and a scattering angle $\zeta = \pi/2+\psi$. $(T_{\rm m}\;T_{\rm a})_1$ and $(T_{\rm m}\;T_{\rm a})_2$ are the total attenuations from the laser facility to the scattering location and from the scattering location to the telescope, respectively. The transition between the Mie domination and the Rayleigh domination is located just a few kilometres above ground level (AGL).}
\label{fig:CLF_geometry}
\end{figure}

Even if LiDARs are now well-known techniques, one of their foundations still raise questions. It has been shown in Sect.~\ref{sec:aerosol_scattering} that aerosols scatter most of the light in the forward direction, with an amplitude depending on the aerosol size. The behaviour of this forward peak is easily understood phenomenologically using e.g.\ the Ramsauer approach~\cite{Ramsauer_PhysScr,Ramsauer_AO}. It has been demonstrated that largest aerosols affect the scattering phase function only in the near-froward peak. On the other side, much less light is backscattered and none of the models, except the Mie scattering theory, is able to explain this backward scattering peak. Nonetheless, it is the latter that is used to probe aerosol population in the atmosphere. This fact is not based on Physics but only on technical considerations: it is easier to install a photo-detector at ground close to the laser facility to record backscattered light that moving it to the top of the atmosphere to record light scattered in the forward direction. Experiments in ultra-high energy cosmic rays have developed the side-scatter technique including a laser facility~\cite{CLF} and a UV telescope to estimate the vertical aerosol optical depth profile $\tau_{\rm a}(h,\lambda)$ in nighttime~\cite{HiRes_CLF,Auger_CLF} (see Fig.~\ref{fig:CLF_geometry}). The main role of the laser facility is to produce calibrated laser ''test beams''. Typically, the beam is directed vertically. When a laser shot is fired, the UV telescope collects a small fraction of the light scattered out of the laser beam. The scattering angles of light from the beam observed by the telescope are in the range of $90^\circ$ to $120^\circ$. Two methods have been developed, both assuming an horizontal uniformity for the molecular and aerosol components. The first method, the so-called Data Normalised Analysis (DNA), is an iterative procedure comparing hourly average light profiles to a reference clear night where light attenuation is dominated by molecular scattering. Using a reference clear night avoids an absolute photometric calibration of the laser. The second method, the so-called Laser Simulation Analysis (LSA), is based on the comparison of measured laser light profiles to profiles simulated with different aerosol attenuation conditions defined using a two-parameters $\{\alpha_{\rm a}(h=0),$ $H_{\rm a}\}$ model~\cite{AtmoHEAD_Valore}. In the latter, the vertical profile for the aerosol component is assumed to be described by a decreasing exponential with an associated scale height $H_{\rm a}$. The corresponding formula for each method is given by
\begin{equation}
\begin{split}
 \tau_{\rm a}(h,\lambda_0)&\stackrel{\rm DNA}{=}\frac{\sin\psi}{1+\sin\psi}\,{\rm ln}\left(\frac{N_{\rm mol}(\psi)}{N_{\rm obs}(\psi)}\right)\\
& \stackrel{\rm LSA}{=}-\frac{H_{\rm a}/\alpha_{\rm a}(h_{\rm gl})}{\sin\psi}\left[\exp\left(-\frac{h - h_{\rm gl}}{H_{\rm a}}\right) - 1 \right],
\end{split}
\label{eq:clf_vaod}
\end{equation}
where $N_{\rm obs}(\psi)$ is the amount of light from the laser beam reaching the detector at the elevation angle $\psi$ and $N_{\rm mol}(\psi)$ is its value in the case of an aerosol-free night. Tests are planned in a R\&D project based in Colorado with a similar UV telescope and a steerable Raman LiDAR replacing the laser facility~\cite{ICRC_Wiencke,AtmoHEAD_Buscemi}. Results will be interesting to crosscheck the validity of each aerosol characterisation technique.

Recently, this side-scatter technique requiring just a steerable laser has been proposed as an end-to-end calibration procedure for the JEM-EUSO telescope on the International Space Station and the imaging air Cherenkov telescopes of the CTA experiment. Whereas a basic central laser facility would be installed for the latter~\cite{CLF_CTA_ICRC13}, it is a worldwide network of ground-based stations called Global Light System (GLS) which is planned for the former~\cite{AdamsJr}. More than 10 stations would be installed and would be operated remotely to generate benchmark optical signatures in the atmosphere with characteristics similar to the optical signals from extensive air showers. Every year, the JEM-EUSO telescope would overfly each station about 300 times with good atmospheric conditions and no Moon. In these two examples, using lasers does not probe the atmospheric conditions but estimate the reconstruction performances of the detector (energy reconstruction, angular reconstruction, trigger efficiency, etc).

\subsection{Cloud cover}
\label{sec:exp_cloud}
Clouds are composed of water droplets or ice crystals attenuating the transmission of optical radiation through the atmosphere. Different techniques can be applied to detect the cloud presence: recording the cloud infrared thermal emission, observing stars in the optical wavelength range, or using LiDARs and detecting backscattered light by clouds. 

\subsubsection{Cloud detection using LiDAR technique}
\label{sec:cloud_lidar}
LiDAR technique described in Sect.~\ref{sec:aerosol_radiative} can be also applied to detect the presence of clouds in the atmosphere. In the same way that aerosol detection, clouds are identified as strong light scatter regions in the backscattered light profiles recorded $S(r)$. A cloud detection algorithm based on the first and second derivative analysis of the signal $S(r)$ permits to retrieve the cloud altitude and the cloud thickness. This method has been, is or is planned to be applied using an elastic-backscatter LiDAR in many ground-based astroparticle physics experiments~\cite{AugerATMON,TaATMON,AtmoHEAD_Toscano}. However, this technique remains poor to distinguish, for instance, between an aerosol layer and a cirrus: both are optically thin and can be at high altitude. A solution would be to measure the shape of scatters since ice particles composing cirrus have a shape much different than aerosols or water droplets. The depolarisation technique could solve this problem: when the emitted laser light is linearly polarised, the backscattered signal recorded can have a different polarisation depending on the shape of scatter centres. Typically, the depolarisation ratio is close to zero in the case of spherical particles, about $25-35\%$ for dust particles, and greater than $40-50\%$ in the case of ice particles. From a technical point of view, it consists in recording the backscattered light in two polarisation channels which are parallel -- co-polarised -- and perpendicular -- cross-polarised -- with respect to the laser polarisation~\cite{Sassen2000,Mattis3003}. To our knowledge, this technique has never been tested in an astroparticle physics experiment, certainly because of the higher complexity compared to elastic-backscatter LiDARs.

Even if the LiDAR technique provides useful information on the spatial extension and the height of the cloud cover, the spatial structure function associated cannot be probed. Nonetheless, this parameter can become important as, for instance, in ground-based astronomical surveys. The two next methods permit to measure it.

\begin{figure*}[!t]
\centering
\resizebox{0.49\textwidth}{!}{%
\includegraphics{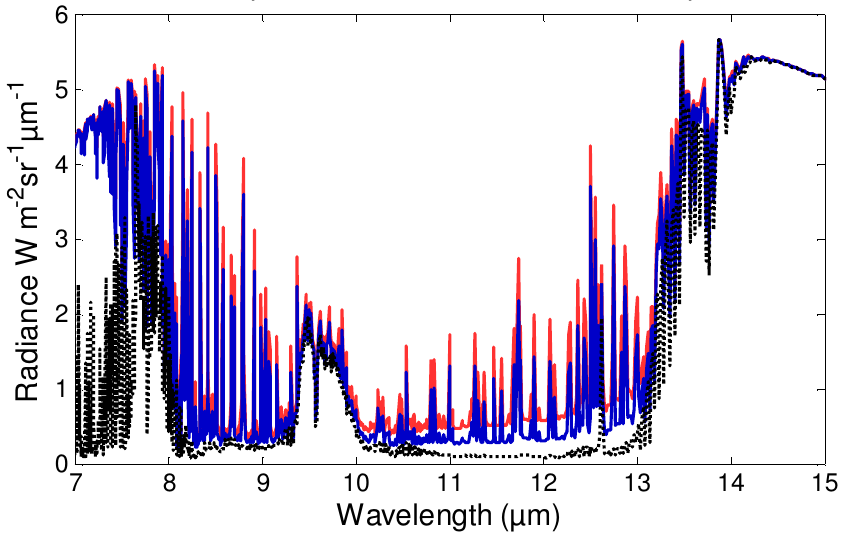}
}
\hfill{}
\resizebox{0.50\textwidth}{!}{%
\includegraphics{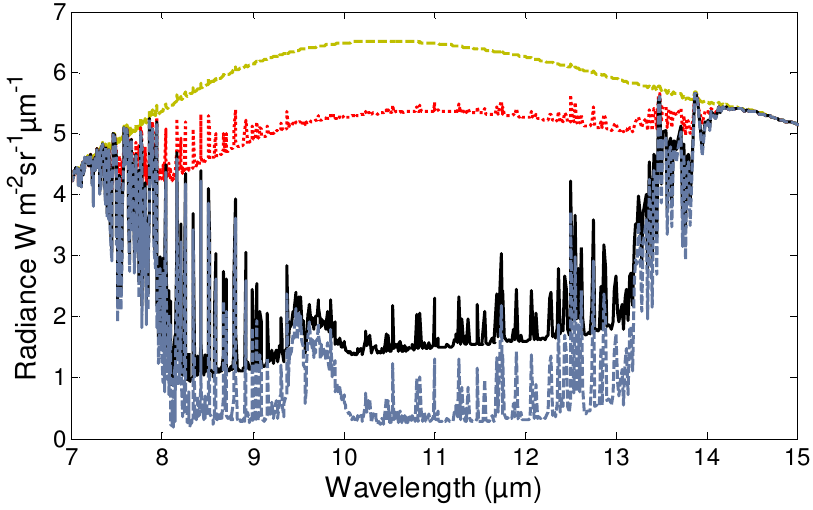}
}
\caption{(left) Simulated infrared spectrum with MODTRAN of a cloud-free atmosphere for different humidity conditions: a dry atmosphere (in black, bottom), a typical atmosphere (in blue, middle) and a wet atmosphere (in red, top). (right) Simulated infrared spectrum with MODTRAN for different types of clouds: cirrus at $10~$km ASL (in black, 2nd from bottom), stratus at $3.9~$km ASL (in red, middle), cumulus at $1.6~$km ASL (in yellow, top) and a cloud-free atmosphere (in grey, bottom) (from~\cite{PWnugent}).}
\label{fig:cloudsIR}
\end{figure*}

\subsubsection{Image analysis of star surveys in optical wavelengths}
Observing stars in the optical wavelength range is one of the two main methods to estimate the cloud cover and its associated spatial structure function. It consists in investigating the presence of stars in the field of view of a camera: using star catalogues available to know their location in the sky and their visual magnitude~\cite{catalog_star1,catalog_star2}, it is possible to measure the atmospheric attenuation between a ground-based camera and the considered star. If a star is not observed, it is deduced that parts of a cloud hide it. Thanks to all-sky CCD cameras, it is possible to observe several hundreds of stars in the same image and to compute the spatial structure function of clouds (assuming the time exposure short compared to the distance travelled by clouds~\cite{GBlanc_private}). Corrections are applied to take into account the decrease of the camera sensitivity with the zenith angle due to the extinction in air masses. Limitations of this method are mainly due to local weather conditions as snow or rain falling on the camera, and the presence of Moon in the field of view saturating the image and increasing the background recorded~\cite{AtmoHEAD_Dusan}. This technique has been applied by the CTA consortium to evaluate the night sky brightness and the cloud fraction (i.e.\ the percentage of the sky covered by clouds) for each CTA candidate site~\cite{ICRC_Dusan}. This setup was capable of detecting a star with visual magnitude up to 6 mag in zenith. In a much efficient way, the same technique can be applied in ground-based astronomical all-sky survey telescopes~\cite{Burke_2} where the higher spatial resolution of the cameras required is balanced by lower limiting star magnitudes probed by these telescopes. 

\begin{figure*}[!t]
\centering
\resizebox{0.47\textwidth}{!}{%
\includegraphics{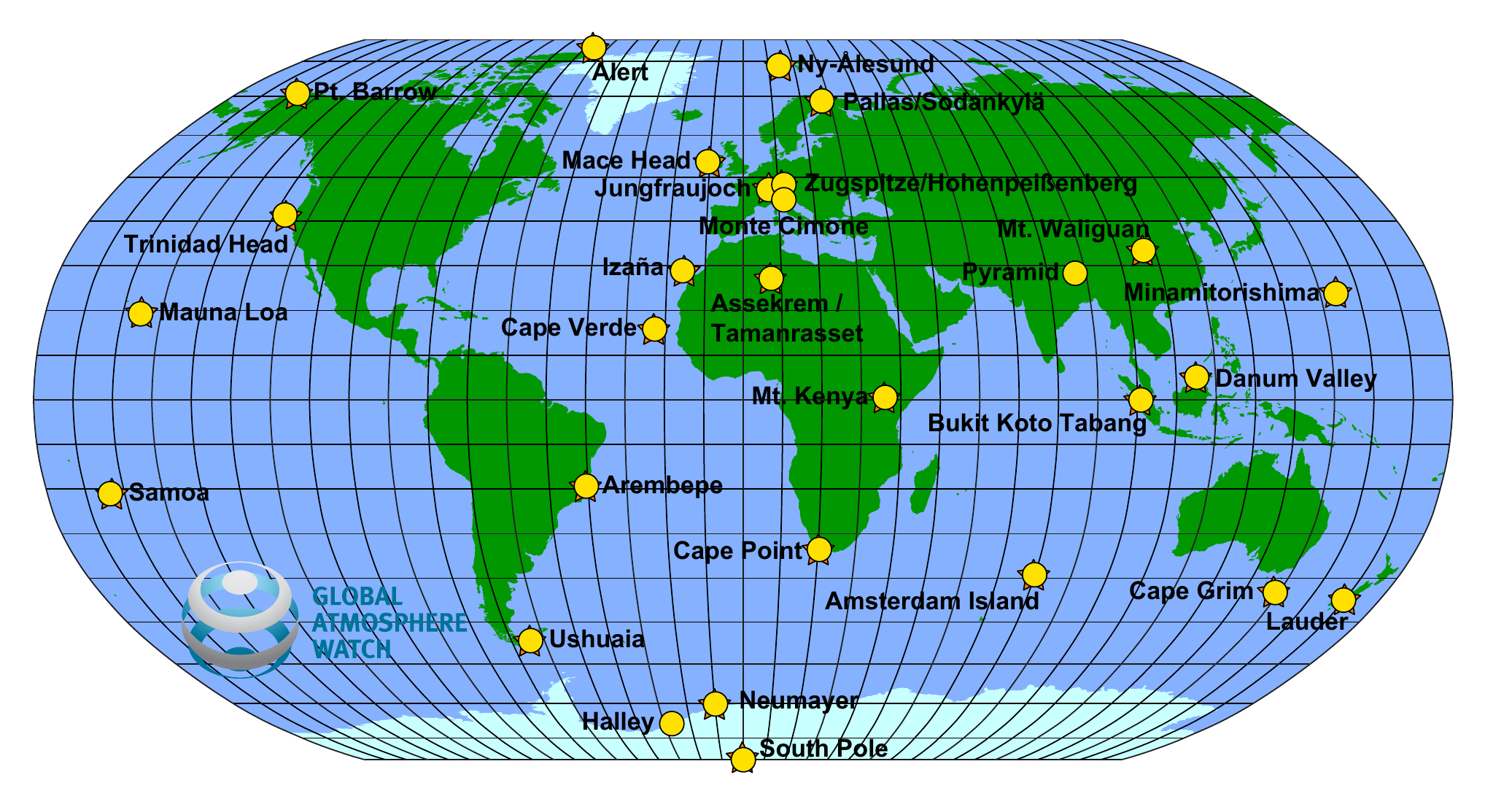}
}
\hfill{}
\resizebox{0.5\textwidth}{!}{%
\includegraphics{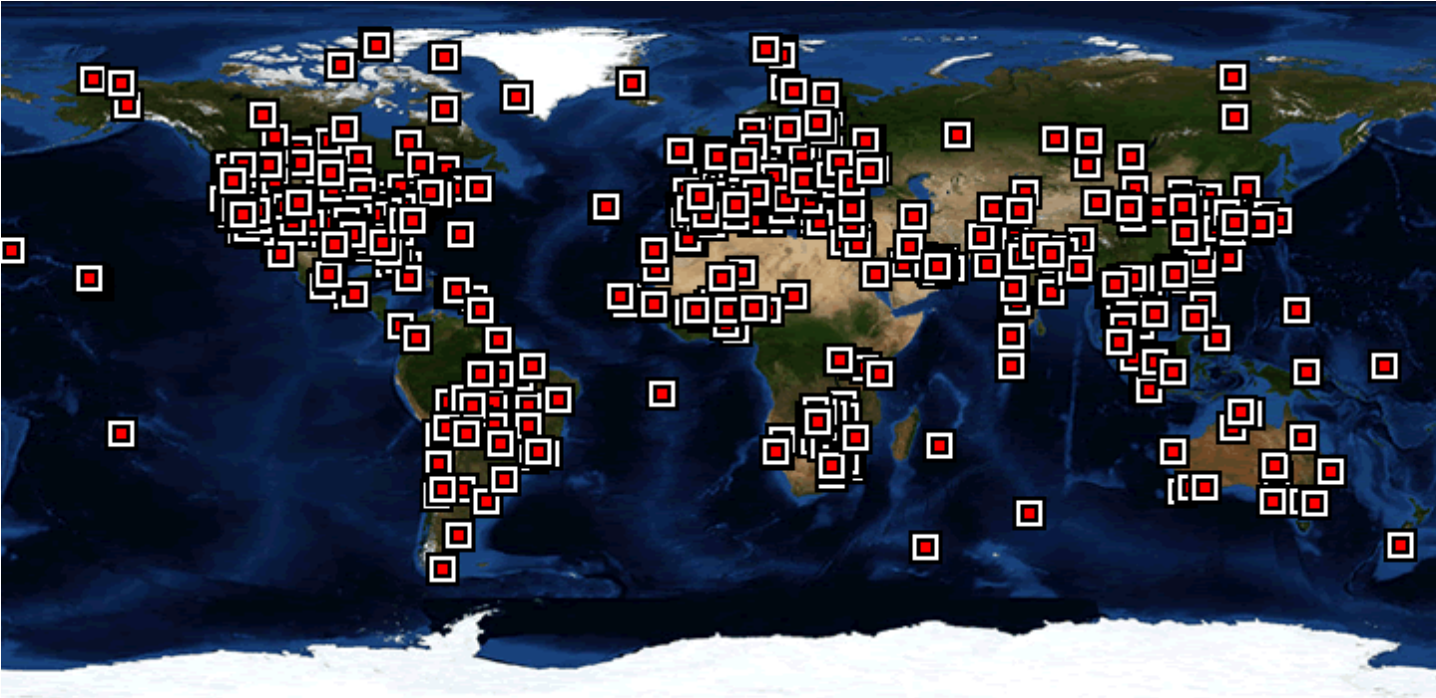}
}
\caption{(left) Network of the 29 ''global'' measurement stations of the GAW programme (map taken from~\cite{GAW_website}). (right) Map representing the geographical distribution of sun-photometers composing the AErosol RObotic NETwork (map taken from~\cite{AERONET_website}).}
\label{fig:networks}
\end{figure*}

\subsubsection{Cloud infrared thermal emission}
Atmospheric emission is the result of infrared emission emitted by certain gases as water vapour, carbon dioxide or ozone when heated by Earth's and solar radiations. Such a radiation is predominant in the infrared $4-50~\upmu$m band. In the case of a cloud-free sky, atmospheric emission can be approximated to a grey or black body. Due to their high water vapour content, clouds radiate also as a black body in the infrared and microwave ranges of the electromagnetic spectrum and their horizontal spatial structure can be measured thanks to their higher effective temperature compared to cloud-free sky. Figure~\ref{fig:cloudsIR}(left) gives the atmospheric emission in cloud-free conditions for a dry atmosphere, a typical atmosphere and an atmosphere in wet conditions. Contrary to wavelengths lower than $7~\upmu$m and greater than about $23~\upmu$m, emission spectrum in this wavelength range does not look like a black body due to absorption or emission by water vapour, carbon dioxide and ozone. In wet atmospheric conditions, the background emission is increased. The long-wave infrared window from $10~\upmu$m to $12.5~\upmu$m is well suited for observing clouds with an upward-viewing system, avoiding bands related to ozone centred at $9.6~\upmu$m and carbon dioxide centred at $14.2~\upmu$m. Figure~\ref{fig:cloudsIR}(right) compares the spectrum emitted by a typical cloud-free atmosphere to the ones produced by three different types of clouds: a cirrus at $10~$km ASL, a stratus at $3.9~$km ASL and a cumulus at $1.6~$km ASL. In presence of clouds, the long-wave atmospheric emission is increased. Cases with stratus or cumulus, i.e.\ optically thick clouds close to the Earth's surface, are very well different with respect to a cloud-free atmosphere in the wavelength range $7-14~\upmu$m, making easy the detection of these clouds by an infrared camera. This is not true anymore for cirrus where the radiometric contrast with the cloud-free atmosphere case is much smaller, mainly due to the fact that cirrus are optically thin and situated at high altitude, increasing the air mass between the observer and the cloud. Because of the emissivity of water vapour, it remains an ambiguity in distinguishing between cirrus and thin fog. One method to avoid this difficulty would be to monitor the water vapour content in the atmosphere in order to suppress its effect during the data analysis. Another technique would be to use at least two filters, one centred on the ozone band at $9.6~\upmu$m -- emissivity independent of the water vapour content -- and a second filter between $10$ and $12.5~\upmu$m. IR cameras recording the cloud infrared thermal emission and associated to a large field of view are usually employed to monitor the cloud cover in astroparticle physics experiments~\cite{KlebeEtAl,Auger_cloud_ICRC13,Auger_cloud_PhDthesis,TA_cloud,JEM_IRcamera,LSST_sebag,RASICAM_Lewis}. Contrary to LiDAR technique, the cloud altitude cannot be estimated directly. However, some algorithms based on the radiance recorded by IR camera on-board a satellite or the International Space Station have been developed by the JEM-EUSO collaboration~\cite{Anzalone_ICRC13} and should be tested soon with the EUSO-Balloon project, a pathfinder of the JEM-EUSO mission~\cite{EUSO_Balloon}. In any other purpose, the Pierre Auger collaboration has checked the agreement between cloud data from satellite and its own measurements of cloud cover using a ground-based laser facility: both methods agree and cloud probability maps covering the region of the observatory are now available every 15 minutes~\cite{Auger_GOES,EJP_JChirinos}.

\subsection{Using data from ground-based atmospheric monitoring networks and satellites}
\label{sec:exp_satellite_networks}
It has been shown in Sect.~\ref{sec:exp_molecular} that it is often better to use data coming directly from global atmospheric models than measuring \emph{in situ} atmospheric state variables by weather radio soundings or ground-based weather stations. This idea is not specific for state variables and can be extrapolated to aerosol data or cloud data provided by ground-based atmospheric monitoring networks or satellites. However, contrary to state variables, precision on these data is not better than measuring aerosol or cloud properties \emph{in situ}. Thus, the usefulness is entirely different here since the main goal is to get a precise knowledge of atmospheric conditions before installing an astroparticle physics experiment on a site and, consequently, to be able to estimate atmospheric effects expected on physics measurements (e.g.\ systematic uncertainties, duty cycle of the detector, etc). This work is for instance currently done for future projects as the JEM-EUSO telescope regarding the cloud cover~\cite{JEM_exposure}, the next imaging air Cherenkov telescope CTA to choose the site candidate offering the best atmospheric conditions~\cite{AtmoHEAD_Vincent,ICRC_CTAcandidateARG1,ICRC_CTAcandidateARG2,ICRC_CTAcandidateUS,ICRC_CTAcandidateSPAIN}, or the LSST telescope to design the best atmospheric monitoring program and to check its corresponding performances~\cite{AlexBoucaud}. To a lesser extent, data from ground-based networks or satellites can be used directly in a real-time atmospheric monitoring program of an astroparticle physics experiment. However, this method is valid only if the distance of the ground-based weather station or the time and space resolutions of the satellite are in adequacy (e.g.\, ozone).  

Ground-based atmospheric monitoring networks represent the first input in global atmospheric models regarding ground-based measurements. Networks impose standardisation of instruments to avoid biases in global data analyses. Whereas stations composing the networks are irregularly dispersed in time and location, global atmospheric models permit to get a well uniform data set. However, if the experiment site is close to one of the elements composing this network, the latter is a wealth of information, much more than a global atmospheric model. The biggest atmospheric network gathering measurements of the chemical composition of the atmosphere is probably the Global Atmospheric Watch (GAW) programme of the World Meteorological Organisation (WMO)~\cite{GAW_website}. Its main goal is to develop a network of measurement stations all around the world. Currently, this network is composed of more than 400 surface-based stations, including 29 elements called ''global stations'' where are operated all the measurements required in the GAW programme (see Fig.~\ref{fig:networks}(left))~\cite{GAWSIS_website}. The atmospheric components monitored by these stations are aerosols, greenhouse gases (e.g.\, carbon dioxide CO$_{\rm 2}$, methane CH$_{\rm 4}$), reactive gases (e.g.\, surface ozone O$_{\rm 3}$, carbon monoxide CO, VOCs, oxidised nitrogen compounds NOx, sulphur dioxide SO$_{\rm 2}$), ozone, etc. In these networks, aerosol data are of great value since aerosols are sampled and analysed chemically, giving a precise characterisation of aerosol properties. Regarding aerosol radiative properties, several global aerosol networks are also available. The most famous is definitely the AErosol RObotic NETwork (AERONET) composed of more than 700 sun-photometers distributed over the whole of continents as depicted in Fig.~\ref{fig:networks}(right)~\cite{AERONET_website,AERONET}. This network, managed by the NASA and the CNRS, monitors the total aerosol optical depth, the precipitable water, but also the aerosol size distribution, the single scattering albedo or the aerosol scattering phase function using inversion algorithms on almucantar scans of solar radiance. Whereas the AERONET provides only properties related to the total aerosol column within the atmosphere, other smaller networks measure the vertical aerosol distribution. The GAW Aerosol LiDAR Observations Network (GALION) is a global aerosol LiDAR network including several sub-networks as the European Aerosol Research LiDAR NETwork (EARLINET)~\cite{EARLINET_website,ACTRIS_website} -- the first aerosol LiDAR network established in Europe in 2000 and which counts currently 27 stations distributed over Europe -- or the Micro-Pulse LiDAR NETwork (MPLNET)~\cite{MPLNET_website} -- a federated network managed by the NASA and counting about 30 micro-pulse LiDAR systems over the world. Regarding clouds, we can cite the CLOUDNET project still in operation in Europe and monitoring in a few sites the cloud coverage and its vertical structure~\cite{CLOUDNET_website}.

Networks of surface-based sensors provide a great amount of data for understanding components of the atmosphere, but they are far from the only source of information. Satellites are also available to offer accurate measurements in regions not covered yet by ground-based weather stations. Satellites are usually divided into two categories: geostationary satellites situated at about $36\,000$ km from the Earth and observing always the same area, and {polar satellites} being on orbit much closer (from $300$ km to $1\,000~$km) and flying a same part of the Earth once to twice a day. Whereas the former are useful to understand time variation of atmospheric quantities, the latter provide a much better spatial resolution of the atmosphere. Satellites embed instruments which can be operated in a ''passive'' mode to detect, for instance, cloud emissivity in infrared or sunlight scattered by aerosols or clouds, or in an ''active'' mode as LiDARs to illuminate the atmosphere and to record backscattered light. Plenty of instruments on-board a satellite have been, are or will be in operation all around the Earth to observe the atmosphere and, consequently, only a non-exhaustive list is given here. We can cite the POLDER satellite~\cite{POLDER}, the PARASOL satellite~\cite{PARASOL}, the CALIPSO satellite~\cite{CALIPSO}, the MODIS instrument on the Terra and Aqua satellites~\cite{MODIS}, the AIRS instrument on the Aqua satellite~\cite{AIRS} or the TOVS instrument on the TIROS satellite~\cite{TOVS} for cloud coverage, water vapour or aerosol optical depth over the Earth (polar), the GOES satellites~\cite{GOES} for cloud coverage on a specific part of the Earth (geostationary), the GOME instrument on the ERS-2 satellite~\cite{GOME} for total ozone, nitrogen dioxide and related cloud information, or the TOMS instrument on Nimbus-7, Meteor-3 and Earth Probe satellites~\cite{TOMS} for total ozone mapping or tropospheric aerosols, etc. Data from satellites are usually available publicly one year later or once the mission is ended and therefore can be used easily to evaluate atmospheric conditions of a site candidate for an astroparticle physics experiment.

\begin{figure*}[!t]
\centering
\resizebox{0.58\textwidth}{!}{%
\includegraphics{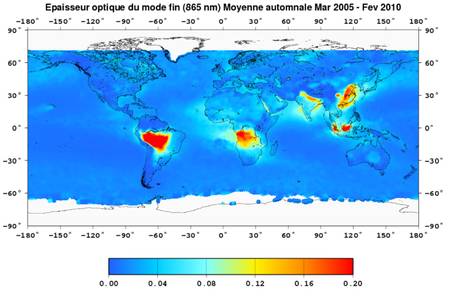}
}
\hfill{}
\resizebox{0.41\textwidth}{!}{%
\includegraphics{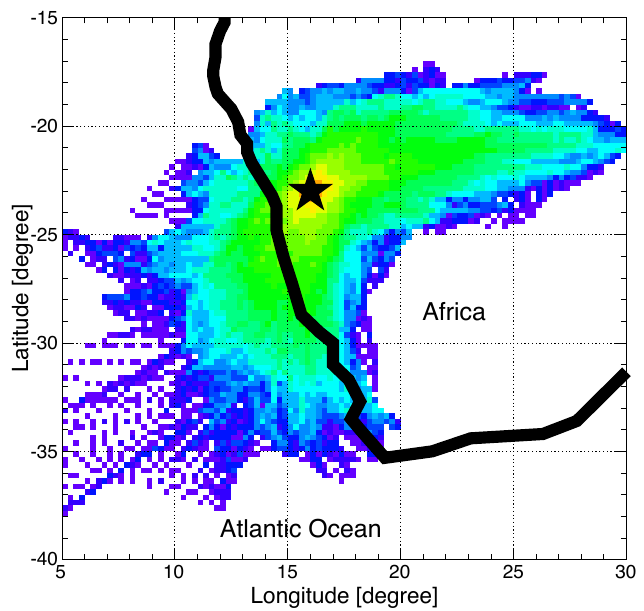}
}
\caption{(left) Aerosol optical depth measured by the Parasol satellite for the months of September, October and November over five years between 2005 and 2009 (from~\cite{Parasol_livre}). (right) Distribution of backward trajectories of air masses at the H.E.S.S. site in 2010 using the HYSPLIT tool. The black star and the black line represent the site location and the South-West African coast, respectively (from~\cite{ECRS_CTA}).}
\label{fig:biomass}
\end{figure*}

\section{Interdisciplinary sciences: the close link between \emph{astroparticles} and \emph{atmosparticles}}
\label{sec:interdisciplinary}
Throughout this review, it has been demonstrated the close link between performance objectives of an astroparticle physics experiment and atmospheric sciences. This is not an isolated case and the same statement could be done with other research fields as geophysics, biology, etc.

\subsection{A growing interest in exotic infrastructures}
Recent years have seen the development of major infrastructures around the Earth in order to considerably increase the competitiveness and the sensitivity of astroparticle physics experiments. This trend has seen the emergence of large international collaborations. Unlike other areas of science where measurements are operated mostly in laboratory, research in astroparticle physics has originality in detection techniques and in infrastructure locations. Although research in astroparticle physics is primarily intended to answer questions in particle physics, astrophysics and cosmology, it has a very close relationship with other research fields through its detection techniques and infrastructures. With this diversity of infrastructures offered in astroparticle physics, unique detectors are available in the world to better understand the Earth, its biodiversity and its environment. The medium in which is located the sensor has its properties varying over time and has to be continuously and precisely monitored. It is this monitoring which offers synergies with Earth sciences. It is in this context that the european agency ASPERA~\cite{ASPERA_web} (now APPEC~\cite{APPEC_web}) organised a conference ''From the Geosphere to the Cosmos'' in December 2010 at the Palais de la D\'ecouverte in Paris (France)~\cite{FromTheGeosphere}. The goal of these two days was to promote and to encourage the development of links between large international collaborations in astroparticle physics and scientists from any other research fields. Initiatives can come directly from the experiments as it is the case with the Pierre Auger collaboration which organised the public workshop IS@AO at Cambridge (UK) in April 2011 to develop interdisciplinary sciences at the Pierre Auger Observatory~\cite{ISATAO}. During this meeting, scientists from a variety of disciplines talked about the potential of the observatory site and exchanged ideas for exploiting it further \cite{EJP_LWiencke}. Among them, we can cite the possible connection between clouds, thunderstorms and cosmic rays~\cite{EJP_Brown}, the observation of elves in the high atmosphere \cite{EJP_RMussa,ICRC13_Tonachini}, or the deployment of a seismic array at the observatory \cite{EJP_Seisme}.

\subsection{An idea to meditate for the next major astroparticle physics projects}
The main idea is that we should promote the link between astroparticle physics and Earth sciences in order to better understand today's measurements to improve the design of future detectors. Recent years have seen the emergence of large international projects in the field of astroparticle physics. Amounts of money involved are so great that we must take full advantage of these new major infrastructures. Also, the level of precision required in these projects is so high that the external environment has to be still better understood and monitored. Even if some embryonic initiatives exist now, we have to think in depth to a collaboration between the two communities. Once a new project in astroparticle physics is planned, the question of an interdisciplinary platform should be asked. The best way would be to contact scientists from other fields and to develop together the best design to complete two aims: {\bf optimising the monitoring of the external environment (land, ocean, atmosphere)}, since challenges in the next major astroparticle physics projects require measurements with ever greater precision, and {\bf developing a multidisciplinary platform}, to host scientists from other research areas to offer them an infrastructure to develop their own studies. Also, a public release of all data related to atmospheric monitoring or oceanographic monitoring should be planned in the experiments. Indeed, they are sometimes situated in areas where weather stations are not numerous leading to a lack of geophysics data in these parts of Earth.

\subsection{Example of the Pierre Auger Observatory and the H.E.S.S. experiment}
Astroparticle physics experiments can play a role in atmospheric sciences too. Using the example of the Pierre Auger Observatory ($35.1^\circ - 35.5^\circ~${S}, $69.0^\circ - 69.6^\circ~${W}, and $1\,300-1\,700~$m ASL, Argentina) and the H.E.S.S. experiment ($23.2^\circ~${S}, $16.3^\circ~${E}, and $1\,800~$m ASL, Namibia), let us see their potential in atmospheric sciences since both projects have accumulated a large database of atmospheric measurements and have developed original techniques to monitor the atmosphere. Without being aware probably, astroparticle physicists have installed these two observatories in similar places for a scientist in atmospheric sciences. Figure~\ref{fig:biomass}(left) gives the map of AOD values for the months of September, October and November over five years obtained by the Parasol satellite. The highest values of aerosol concentration during Austral spring are found in China and India because of urban pollution and industry, and in Indonesia, central Africa and Amazonia because of the phenomenon of biomass burning. It is now well-known that wildfire emissions in these regions, occurring mainly during the dry season, strongly affect a vast part of the atmosphere in South Hemisphere via long-term transportation of air masses~\cite{Andreae,Fiebig,Freitas,Generoso}. The impact of emissions from fires on global atmospheric chemistry, and on atmospheric burden of greenhouse gases and aerosols are recognised even if it remains to quantify it \cite{Keywood}. To illustrate this purpose, Figure~\ref{fig:biomass}(right) gives the distribution of backward trajectories of air masses at the H.E.S.S. site~\cite{ECRS_CTA}. It has been obtained using the HYSPLIT tool \cite{HYSPLIT_1,HYSPLIT_2}, a well-known air-modelling programme in atmospheric sciences for calculating air mass displacements from one region to another. A part of air masses comes directly from the Northern region of Namibia, typically shen biomass burning is observed. This assumption done studying the air mass origin was confirmed later by measurements of aerosol optical depth at the H.E.S.S. site~\cite{Hahn_biomass}.

Because the South Hemisphere is mainly constituted of oceans, the only possible dust (i.e.\ atmospheric mineral aerosols) sources are, Argentina, South Africa and Australia, making another similitude for the two astroparticle physics experiments. Atmospheric dust is one of the major vectors feeding open ocean surface waters with trace metals~\cite{RL_1}. Even at extremely low concentrations, trace metals are micro nutrients necessary for the growth of phytoplankton~\cite{RL_2}. By this way, the trace metals are linked to climate since they affect the capability of the marine biomass to trap CO$_2$~\cite{RL_3}. The Austral region ranging from about $40^\circ$ and $65^\circ$ S is one of the major CO$_2$ sink. This region is also very remote from continents and thus atmospheric dust exhibits very low concentrations~\cite{RL_4,RL_5,RL_6,RL_7}. This oceanic area is a HNLC region (High-Nutrients Low-Chlorophyll) and dust deposition could be a severe limiting factor for the primary production~\cite{RL_8,RL_9}. Given the key role of the Austral Ocean on global climate, scientists in atmosphere sciences have initiated studies characterising mineral aerosols in Patagonia and more generally in South America since Argentina is suspected to be the major dust source for the oceanic region ranging between $40^\circ$~S and $60^\circ$~S~\cite{Losno_private,EJP_KLouedec}.

\section{Summary and conclusion}
\label{sec:summary}
Astroparticle physics experiments require still greater precision in measurements to answer questions in particle physics, astrophysics and cosmology. Some of these experiments use the atmosphere as a part of their detector. In order to reduce as much as possible systematic uncertainties related to the atmosphere, extensive atmospheric monitoring programs have been, are or will be developed by collaborations. It has been shown that all the astroparticle physics experiments are not at the same stage in atmospheric monitoring: whereas collaborations in ultra-high energy cosmic rays use already techniques developed in atmospheric sciences to probe atmospheric properties and correct its effect in their measurements, scientists in very-high energy gamma rays or ground-based astronomical surveys are still in a stage where atmospheric measurements are used only as a quality cut on data selection. However, this would not be true anymore in the next major projects where the challenge of environment monitoring will be a key element for developing instrumentation. In order to better carry out these future projects, it makes sense to collaborate with scientists in Earth sciences to choose the best methods and techniques to reach scientific goals.

Concerning atmospheric measurements, it has been shown that many instruments and techniques developed in atmospheric sciences are available. Depending on the atmospheric component monitored -- molecular, aerosol or cloud -- same instruments will not be used. Before installing any instrument on site, it is necessary to know the effect of the chemical component planned to be measured in the wavelength range studied and its time and spatial variations. Since in the near future astroparticle physics experiments will require extensive atmospheric monitoring programs with many instruments, the idea to join a worldwide atmospheric network has to be considered. Indeed, experiments are sometimes situated in places with a few weather stations available and joining such networks could represent an opportunity for both research fields.

\section*{Acknowledgments}
The author would like to thank his colleagues of the Pierre Auger collaboration, and especially L. Wiencke, B. Keilhauer, B. Dawson, V. Verzi, M. Will, M.I. Micheletti and M. Unger for fruitful discussions during these last six years. Also, K.L. thanks M. Urban for having given him the opportunity to integrate the atmospheric monitoring task inside the Pierre Auger collaboration during his PhD thesis. Finally, K.L. would like to mention encounters having inspired him these last years with D. Veberi\v{c}, R. Losno and S. BenZvi.  
~\\

%
%


\end{document}